\documentclass[preprint2]{aastex}
\usepackage{graphicx}
\usepackage{natbib}
\usepackage{amsmath}
\usepackage{amssymb}
\usepackage{rotating}

\newcommand{\Unit}[1]{\ensuremath{\mathrm{~#1}}} 

\newcommand{\yr}{\Unit{yr}}
\newcommand{\Myr}{\Unit{Myr}}
\newcommand{\Gyr}{\Unit{Gyr}}

\newcommand{\kpc}{\Unit{kpc}}
\newcommand{\Mpc}{\Unit{Mpc}}

\newcommand{\msun}{\Unit{M}_{\odot}}

\newcommand{\halpha}{{H}{$\alpha$}\ }
\begin{document}

\shortauthors{Fouesneau et al.}  
\lefthead{\textsc{Fouesneau et al.}}
\righthead{\textsc{M~83 star cluster populations with stochastic models}}

\title{Analysing star cluster populations with stochastic models: the HST/WFC3
sample of clusters in M\,83}
\author{ Morgan Fouesneau\altaffilmark{1,2},
         Ariane Lan{\c c}on\altaffilmark{1},
	 Rupali Chandar\altaffilmark{3},
	 Bradley C. Whitmore \altaffilmark{4}
       } 
\altaffiltext{1}{Observatoire astronomique and CNRS UMR 7550, Universit\'e de Strasbourg, Strasbourg, France}
\altaffiltext{2}{Department of Astronomy, University of Washington, Seattle, Washington, USA}
\altaffiltext{3}{Dept. of Physics and Astronomy, University of Toledo, Toledo, Ohio, USA}
\altaffiltext{4}{Space Telescope Science Institute, Baltimore, Maryland, USA}

\email{morgan.fouesneau@astro.u-strasbg.fr}

\begin{abstract} 
{ The majority of clusters in the Universe have masses well below
$10^5\msun$.  Hence their integrated fluxes and colors can be affected by the
presence or absence of a few bright stars introduced by stochastic sampling of
the stellar mass function.}  
{Specific methods are being developed to extend the analysis of cluster energy
distributions into the low-mass regime. In this paper, we apply such a method to
real observations of star clusters, in the nearby spiral galaxy M\,83.}
{We reassess the ages and masses of a sample of $1242$ clusters for which
UBVI\halpha fluxes were obtained from observations with the WFC3 instrument on
board of the Hubble Space Telescope.  Synthetic clusters with known properties
are used to characterize the limitations of the method (valid range and
resolution in age and mass, method artifacts).}
{ The ensemble of color predictions of the discrete cluster models are in good
agreement with the distribution of observed colors.  We emphasize the important
role of the \halpha data in the assessment of the fraction of young objects,
particularly in breaking the age-extinction degeneracy that hampers an analysis
based on UBVI data only. We find the mass distribution of the cluster sample to
follow a power-law of index $-2.1 \pm 0.2$, and the distribution of ages a
power-law of index $-1.0 \pm 0.2$ for $\log(M/\msun)>3.5$ and ages between
$10^7$ and $10^9\yr$. An extension of our main method, that makes full use of
the probability distributions of age and mass obtained for the individual
clusters of the sample, is explored.  It produces similar power-law slopes and
will deserve further investigation.}
{ Although the properties derived for individual clusters significantly differ
from those obtained with traditional, non-stochastic models in about 30\% of the
objects, the first order aspect of the age and mass distributions are similar to
those obtained previously for this M\,83 sample in the range of overlap of the
studies. We extend the power-law description to lower masses with better mass
and age resolution and without most of the artifacts produced by the classical
method.}
\end{abstract} 
\keywords{Galaxies: Individual (M\,83), Star clusters --- Methods: data analysis
--- Techniques: photometric --- Method: data analysis, statistical}

\maketitle

\section{Introduction} 

Many and possibly most stars form in clusters rather than individually. Much
observational effort has therefore been invested to determine the distributions
of star cluster ages and masses \citep{Searle1980, Larsen2000, Billett2002,
Hunter2003, Fall2005, Dowell2008, deGrijs2009, Larsen2009, Chandar2010,
Bastian2011}, and to determine the dominant mechanisms of cluster formation and
disruption \citep{Kroupa2002, Lamers2005, Whitmore2007, Parmentier2008,
Fall2009, Elmegreen2010, Converse2011}. With the Hubble Space Telescope and
especially its latest survey instrument WFC3, cluster samples have been detected
in external galaxies down to a regime in which cluster luminosities overlap
those of individual bright stars, and hence to lower masses \citep{Johnson2011}.
Because cluster luminosity (and mass) distributions rise steeply towards lower
luminosities (and masses), these deeper surveys have led to significantly larger
cluster samples.

In the low-mass regime, clusters of a given age and mass are predicted to
display a broad range of integrated luminosities and colors, mostly as a
consequence of the random sampling of the upper part of the stellar mass
function \citep{Barbaro1977, Girardi1993, Lancon2000, Bruzual2002, Cervino2006,
Deveikis2008}.  For the sake of simplicity, we will refer to stellar
population model implementations that explicitly predict luminosity and color
{\em distributions} as {\em stochastic models}. The
predicted distributions depend in non-trivial ways on the cluster mass.
Stochastic models are usually based on Monte Carlo simulations, although first
attempts at analytical approaches have been made \citep{Cervino2006}.
Traditionally, stellar population models have been implemented with the
assumption of a fully sampled initial stellar mass function (IMF): each stellar
mass bin along an isochrone contains exactly the mass fraction prescribed by the
IMF, regardless of the fact that this may lead to unphysical non-integer numbers
of stars in some of the most luminous phases of evolution. These
implementations, which we will refer to as {\em continuous models}, do not
predict distributions of cluster luminosities and colors but rather their mean
values (and sometimes their first moments). 

Monte-Carlo simulations of low-mass clusters have shown that the range of
broad-band colors predicted with this type of stochastic model are in
rather good agreement with observed ones in the Milky Way and in the Magellanic
Clouds \citep{Girardi1993, Piskunov2009, Popescu2010b}. The color and
luminosity distributions are complex in the low-mass regime.  They display
multiple peaks. The most probable luminosities and colors differ from the mean
color of the distribution, and conversely the mean properties are unlikely to
occur in any actual cluster. Therefore ages and masses derived from the analysis
of cluster luminosities and colors are expected to depend on the type of model
used (i.e. stochastic or continuous). For different combinations of broad-band
filters, \citet{Maiz2009} and \citet{Fouesneau2010} confirmed that the
analysis of the integrated light of {\it synthetic} low-mass clusters with {\it
continuous} models leads to biases in the assignment of ages to individual
objects.

Assigning fundamental properties to individual clusters and studying the global
properties of the cluster population of a galaxy are two different issues. For
instance, \citet{Fouesneau2010} showed that the mass estimates determined for
low-mass clusters with continuous models would be dispersed around the
real values, but that there would not be a strong global impact on the cluster
mass distribution itself. This work also indicated that high resolution features
in the cluster age distribution would be affected, but \citet{Fouesneau2010} did
not examine whether any of these biases would be strong enough to modify the
global, low resolution age distribution of a typical cluster sample.

The main purpose of the article of  \citet{Fouesneau2010} was to introduce 
two analysis methods for the interpretation of integrated fluxes of low-mass
star clusters, both based on a large collection of Monte-Carlo simulations
of clusters of finite numbers of stars, i.e.  on stochastic models.  One
analysis method is a traditional $\chi^2$ minimization.  The second one rests on
the calculation of posterior probability distributions in age-mass-extinction
space. The latter refers explicitly to Bayes' theorem, in which the posterior
probabilities of model parameters given a set of data are expressed in terms of
their prior distributions and of the probability distributions of observational
errors.  Although the results from the analysis methods have their own
model-dependencies through the adopted population synthesis code and the
priors, it is clearly a conceptual improvement to move away from the restricted
framework of continuous population synthesis models.  Studies in the regime well
below $10^4\msun$, where the literature on the subject does not usually venture,
become possible. This should allow us to examine cluster formation and
disruption histories in more detail in the future.

In this paper, we apply the methods of \citet{Fouesneau2010} to the WFC3 sample
of star clusters in M\,83 first presented by \citet{Chandar2010}. Messier 83,
the ``Southern Pinwheel'' galaxy, is one of the best nearby analogs of the Milky
Way.  These data allow us to compare standard age and mass estimates based
on continuous models with those based on stochastic ones for a sample of {\em
real}\ objects. We also investigate how the use of stochastic models
affects the age and mass distributions of the sample as a whole. 

The rest of this paper is organized as follows.  Section~2 summarizes the data,
cluster selection and photometry, and Section~3 summarizes and extends
predictions from the stochastic models developed by \citet{Fouesneau2010}.
Section~4 compares the luminosity and color distributions of the colors and
model predictions.  Expectations from artificial clusters are presented in
Section~5, and results for real clusters in M\,83 are presented in Section~6.
Section~7 discusses several key issues, including the importance of using a
narrow-band filter in the analysis, and a direct comparison of results from
stochastic and continuous models. More details relevant to this comparison are
provided in two appendices.  Section~7 also present preliminary exploration
of an alternative determination of the age-mass distribution of the cluster
sample of M\,83, one which uses the probability distributions of individual
clusters rather than just their most probable values.  Section~8 presents the
main conclusions of our work. 

\section{Data and Observations}

\begin{figure*}
        \begin{center}
                \includegraphics[width=2\columnwidth]{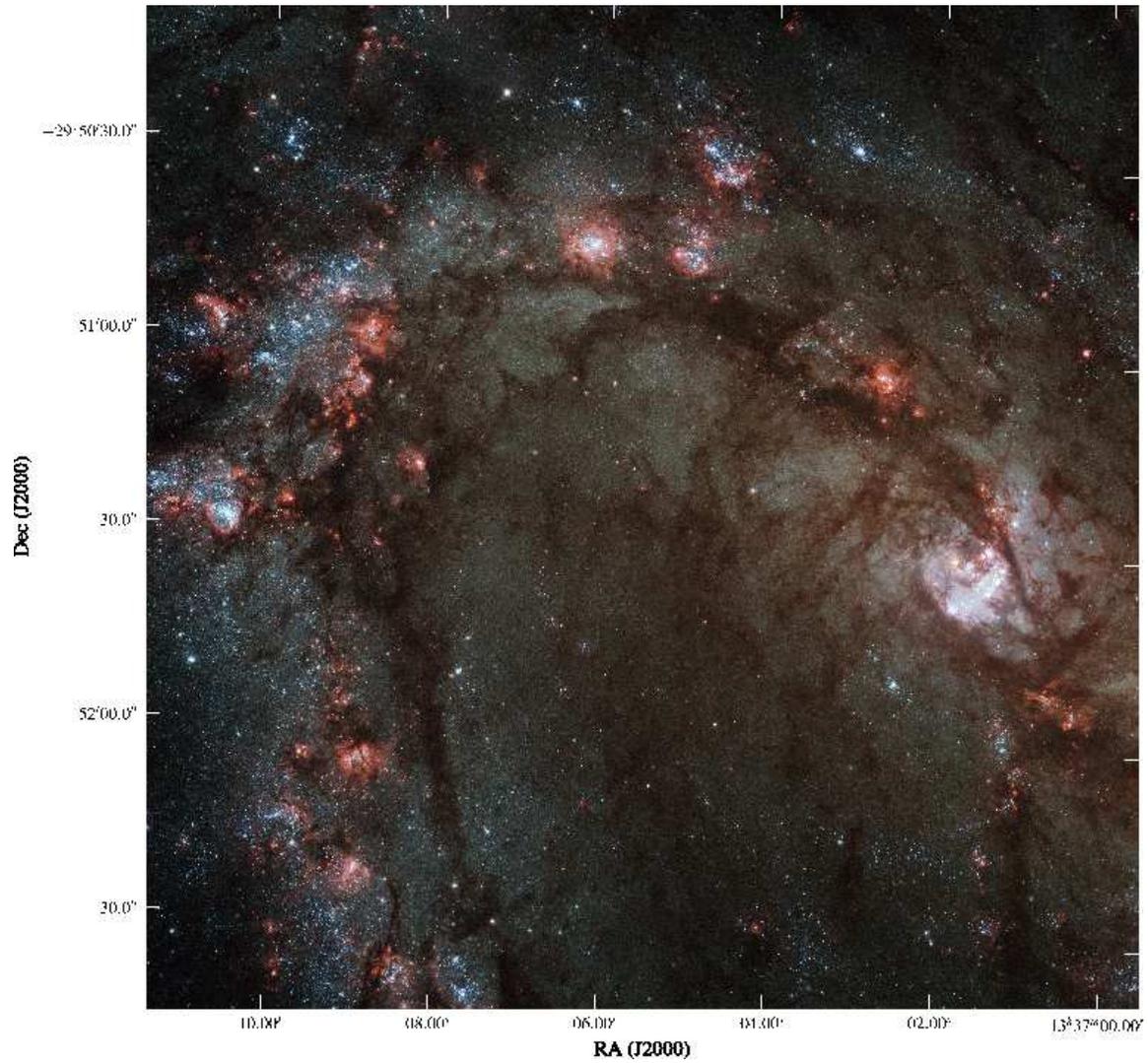}
        \end{center}
        \caption[Color image of the observed region of M\,83]{Color image of the
        observed region of M\,83 with the WFC3 instrument on-board HST.  The
        B-band (F438W) image is shown in blue, the V-band (F555W) image in green
        and a combination of the I-band (F814W) and H\,$\alpha$ in red.
        The covered area represents $2.75\arcmin \times 2.75\arcmin $ ($\approx
        3.6\times3.6\,\kpc^2$).
        \label{fig:m83image}}
\end{figure*}

In order to take advantage of the wavelength coverage of WFC3, images of M\,83
were taken as part of the ERS1 program 11360 (PI: O'Connell) through seven
broad-band filters from the UV to the near infrared: F225W (UV), F336W (U),
F438W (B), F555W (V), F814W (I), F110W (J), F160W (H).  They cover the nucleus
and the North-Eastern part of the galaxy.  The color image in
Fig.\,~\ref{fig:m83image} illustrates the observed region. In addition, narrow
band filter observations were taken for the following emission lines:
$[$\ion{O}{3}$]$ (F373N), H$\beta$ (F487N), \ion{O}{2} (F502N),
H$\alpha$\,+\,[\ion{N}{2}] (F657N), $[$\ion{S}{2}$]$ (F673N). 

In this paper, we focus only on the broad band UBVI photometry (HST/WFC3
passbands) and the narrow band \halpha measurements (F657N).  The sample
contains $1242$ star clusters with measurements in these five bands\footnote{
The cluster catalog, incl. photometry and positions, can be retrieved at:
\url{http://archive.stsci.edu/prepds/wfc3ers}}.  They were selected based on a
``white" light image (i.e. a co-added, rms-normalized combination of the $U$,
$B$, $V$ and $I$ filters), in order to give roughly equal weight to all
wavelengths. A combination of size and concentration criteria 
measured in the V band was used to separate individual 
stars from clusters, as described in \citet{Chandar2010}.  
For each object, circular aperture
photometry with an aperture radius of $3$ pixels ($0.0396$\arcsec\,pix$^{-1}$)
produced integrated flux values which were then corrected for foreground
galactic extinction \citep[appendix B]{Schlegel1998}.  Measurement uncertainties
are typically spread between $0.05$ and $0.25$\,mag for the { B, V and I} broad
bands (cf. Sect.\,\ref{sec:resolution}).  About one half of the objects have
uncertainties larger than $0.25$ in U and \halpha filters. The average errors
increase non linearly with magnitude.  Instrumental magnitudes were
converted into the VEGA magnitude system by applying the following zero-points:
F336W=23.46, F438W=24.98, F555W=25.81, F814W=24.67, F657N=22.35. We assume a
distance of $4.5\Mpc$ to M\,83 as found in \citet{Thim2003}, which corresponds
to a distance modulus of $m-M = 28.29$. 

\section{Population synthesis models} \label{sec:models}

We aim at studying the age-mass distribution of the clusters in M\,83, using the
method of \citet{Fouesneau2010}.  The method is based on a large collection of
Monte-Carlo simulations of individual clusters.  The synthetic clusters are
constructed with the population synthesis code {\sc P\'egase.2n} (Fouesneau et
al. in prep.), which is derived from {\sc P\'egase} \citep{Fioc1997}.  As in the
original (continuous) population synthesis code, the underlying stellar
evolution tracks are those of the Padova group \citep{Bressan1993}, with a
simple extension through the thermally pulsating AGB based on the prescriptions
of \citet{Groenewegen1993}. The input stellar spectra are taken from the library
of \citet{Lejeune1997}.  The stellar Initial Mass Function (IMF) is taken from
\citet{Kroupa1993}, and extends from $0.1$ to $120\msun$. Nebular emission
(lines and continuum) is included in the calculated spectra and broad band
fluxes under the assumption that no ionizing photon escapes. Line ratios
are computed as in \citet{Fioc1997}. When extinction corrections are
considered, they are based on the standard law of \citet{Cardelli1989}. 
The synthetic photometry for the artificial clusters is computed using the
response curves of the HST/WFC3 filters. A reference spectrum of Vega provides
zero magnitude fluxes \citep{Bohlin2007}.

In the context of studying M\,83, the model collection is restricted to solar
metallicity, the metallicity of this galaxy according to \citet{GildePaz2007}.
The simulated cluster set has been extended to higher masses than available
in \citet{Fouesneau2010}, and now covers masses from $10^3$ to
$5\times10^5\msun$ and ages from $1\Myr$ to $20\Gyr$.  With $1.4 \times 10^7$
individual objects, the collection  is large enough to include all reasonably
likely cluster properties.
The ages of the synthetic clusters are drawn from  a power law
distribution with index $-1$ (equal numbers of star clusters per logarithmic
age bin), rounded to integer multiples of $10^6$\,yr.  For practical reasons
(the need to include more massive clusters without having to recompute
prohibitive numbers of low-mass clusters), the mass distribution in the new
collection of models also follows a power law of index $-1$, instead of the
previously adopted $-2$ in  \citet{Fouesneau2010}\footnote{The number of
massive clusters in the current collection is larger than strictly necessary.
The choice of the power law index can be optimized in the future to construct
smaller representative cluster sets.}. 

The age and mass distributions of the synthetic clusters, together with the
values allowed for extinction, are the main assumptions (priors) of the
inversion. These distributions account for two major qualitative trends
found in star-forming galaxies: low mass clusters are more numerous than high
mass clusters, and, because of a variety of efficient disruption mechanisms,
young clusters are more numerous than old ones. In the future, it will be
interesting to implement an iterative inversion, in which the derived cluster
age and mass distributions, corrected for selection effects, are re-injected as
priors until the procedure converges.

\begin{figure*}
  \begin{center}
	\includegraphics[width=1.9\columnwidth, clip=]{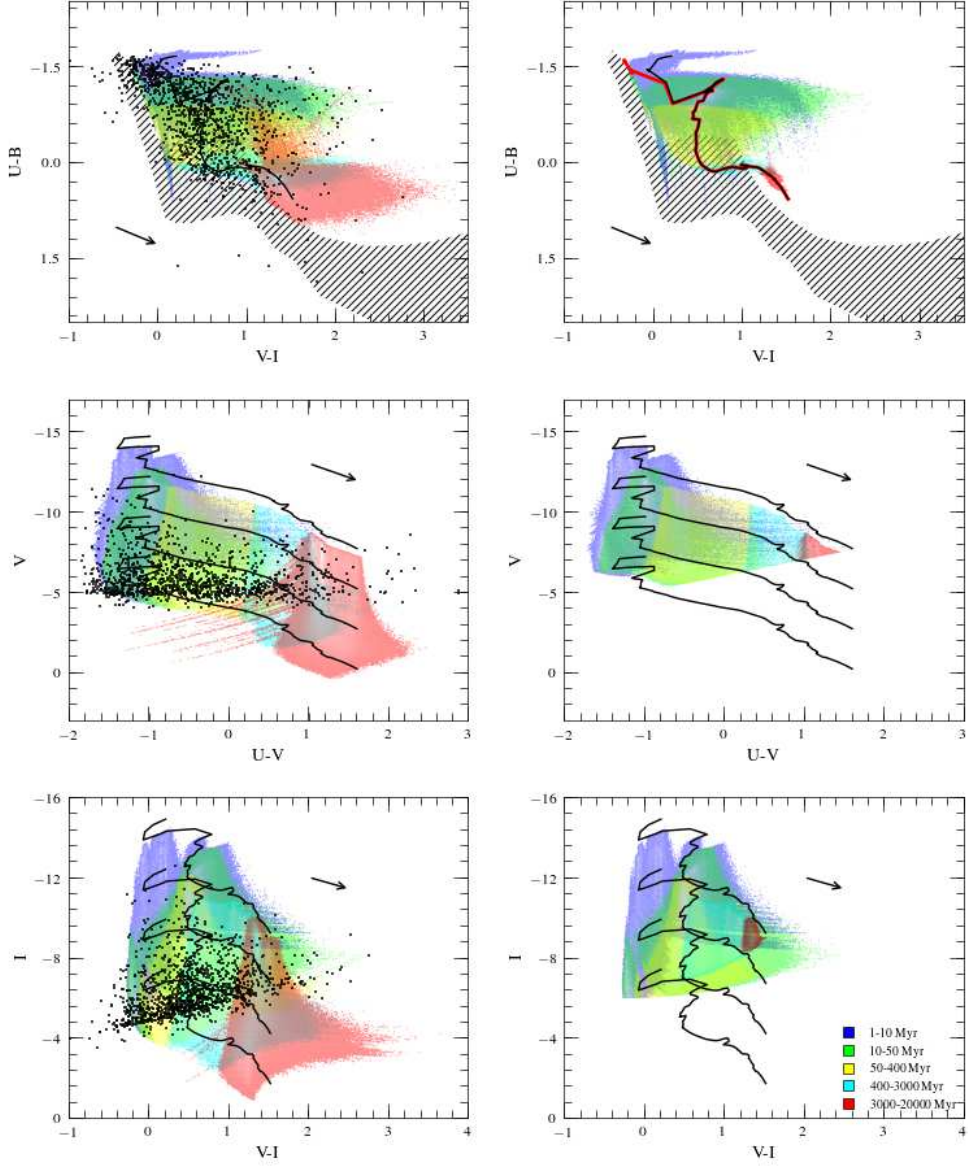}
   \end{center}
   \caption[Color-color and color-magnitude diagrams of the sources]
   {Color-color and color-magnitude diagrams of the observed clusters (black
   points) and the models.  Magnitudes are in the Vega system, in the
   WFC3 filters.  Solid lines show the predictions obtained with the {
   corresponding} ``continuous'' population synthesis models for
   $10^3$, $10^4$, $10^5$, and $10^6\msun$.  The red line in the 
   top right panel shows the effect of excluding any ionized gas emission
   from the predicted fluxes. Colored dots represent the
   discrete clusters available in the Monte-Carlo collection. The
   color-code is given in the bottom right panel. In the right hand
   panels, only discrete models brighter than the observational limits in
   UBVI+\halpha are shown. Models are not reddened in any panel,
   but the extinction vector for $A_V=1$ is shown. Hatched regions
   represent the expected colors of individual stars.
   \label{fig:complete_cmds}}
\end{figure*}

\section{Luminosity and color distributions
\label{sec:colors_distrib}}

Previous work by \citet{Chandar2010} shows that the V band luminosity function
of the observed clusters can be described globally as a power-law with index
$\sim -2$ down to the completeness limit located near $M_V=-5$. For the four
other bands used here, the turn-over points in the luminosity distributions are
located near $M_U=-6$, $M_B=-5$, $M_I=-6$ and $M_{H_\alpha}=-6$ (in the absolute
Vega magnitude system).  For brevity, we will refer to these turn-over points as
``completeness limits''.  {Note that \citet{Chandar2010} estimated that this
cluster catalog contains contamination at the approximately 15\% level from
individual stars in crowded regions and from chance superpositions of a few
stars.}

Figure~\ref{fig:complete_cmds} compares the loci of the observations and the
models in three projections of color-magnitude space.  In the left hand panels,
the data are shown together with the complete set of models, which will be used
to assign age, mass and extinction estimates to each individual object. The
majority of the observed clusters lie well within the regions covered by the
synthetic clusters. It is also clear that some of the observations cannot be
reproduced with continuous population synthesis models, even when allowing for
extinction. The clusters with the bluest (V-I) colors are those whose post-main
sequence happens to be underpopulated, while the clusters with the reddest (V-I)
colors may either happen to have more luminous red stars than average, or be
severely reddened (or both). 

At very young ages the color predictions are sensitive to the prescription
adopted for the nebular emission: the nebular fluxes added to the stellar fluxes
produces redder WFC3 V$-$I colors (to allow comparison, the red line in
Fig.~\ref{fig:complete_cmds} illustrates the effect of discarding the gas
contribution, in the case of continuous models). This leads to the hook-like
extension seen in blue at the top of the upper panel of
Fig.~\ref{fig:complete_cmds}, composed of low-mass clusters (a few $10^3\msun$)
with a number of ionising stars in excess of the average.  They do not have
counterparts in the observed cluster sample.  The models in that part of the
diagram represent only $0.1$\,\% of the models with ages below $3\Myr$, and will
therefore have little effect on any of our results. On the other hand, the
very blue colors observed for a subset of obviously young clusters in M\,83
suggest that the nebular emission included in our models is too high for some
objects; this could affect age-dating, mainly between $1$ and $4\Myr$.

One can conclude from the two bottom-left panels of Fig.~\ref{fig:complete_cmds}
that a majority of the observed clusters have masses lower than $10^4\msun$.  At
these masses, the predicted optical fluxes of star clusters are spread quite
widely. For instance, the 90\,\% confidence intervals for the predicted
fluxes respectively have widths of $0.3$\,dex at $5\Myr$ in the V band,
$0.1$\,dex at $50\Myr$ in V, and $0.5$\,dex at $5\Myr$ in the \halpha filter
(based on our Monte Carlo simulations). The distributions have complex shapes.
This is the mass regime where the use of the method of \citet{Fouesneau2010} may
be expected to produce the most significant changes with respect to traditional
approaches. A main purpose of this paper is to illustrate to what extent this
does (or does not) modify results for individual clusters and for the 
sample as a whole.

The right hand panels illustrate some of the complex effects of magnitude
limits.  The models shown are those brighter than the ``completeness limits'' of
the observations in all five bands. {Some clusters are observed below these
limits, because the luminosity distributions of the observations do not fall off
abruptly beyond their turn-over: $10$\% of the clusters are below the turn-over
of the luminosity function in $V$ or H$\alpha$, $30$ to $40$\% in $U$, $B$, and
$I$. } More complex figures are obtained when reddened models are included (not
shown).  In the stochastic context, a magnitude cut does not simply reject {\em
all} the clusters of a given mass that have reached a critical age.  On the
contrary the mass-to-light ratios in this regime have a large range of possible
values.  At a given mass and age only those stochastic clusters that happen to
be fainter than the magnitude limits will be rejected, while those that happen
to be brighter will be kept. From the figures shown here one should retain that
the effects of a stochastically sampled IMF will be more important for the
subset of young clusters than for the old ones, because the latter must be of
high mass to be detected.

The model distributions are not and should not be truncated in magnitude when
used to assign fundamental properties to individual clusters.  The presence
of low luminosity clusters in the Monte-Carlo sample can do no harm in the
analysis of a cluster observation.  On the other hand, the empirical sample may
contain a small number of clusters with intrinsic luminosities below the
``completeness limits'', that made it into the sample only thanks to favourable
error bars.  With a truncated theoretical distribution, the analysis would fail
to provide correct properties for such objects.

In the future however, it will be worthwhile to construct more
sophisticated versions of the truncated model distributions in the right hand
panels of Fig.\,\ref{fig:complete_cmds}.  If the original Monte Carlo collection
can be modified to account for all observational effects present in the M\,83
sample, the resulting synthetic color-magnitude distributions can be compared
directly with the observed ones.  This approach would be a direct analog of what
has slowly become common practice in the analysis of the resolved {\em stellar}
color-magnitude diagrams of nearby galaxies. In the latter case the aim is to
determine star formation histories. In our case current age-mass distributions
could be estimated, without the detour through the individual analysis of each
observed cluster.  We have not yet implemented the tools necessary for such a
study.  The observational effects relevant to a cluster sample are more
complex than those one deals within star samples (i.e. variable cluster surface
brightness, size, effect of crowding, background fluctuations\dots). Correcting
for them lies beyond the scope of this paper, and we will rather restrict
conclusions about the cluster population of M83 to ranges where incompleteness
is not severe.

\section{Expectations from artificial data}
\label{sec:resolution}

\begin{figure}
        \begin{center}
	  \includegraphics[width=\columnwidth]{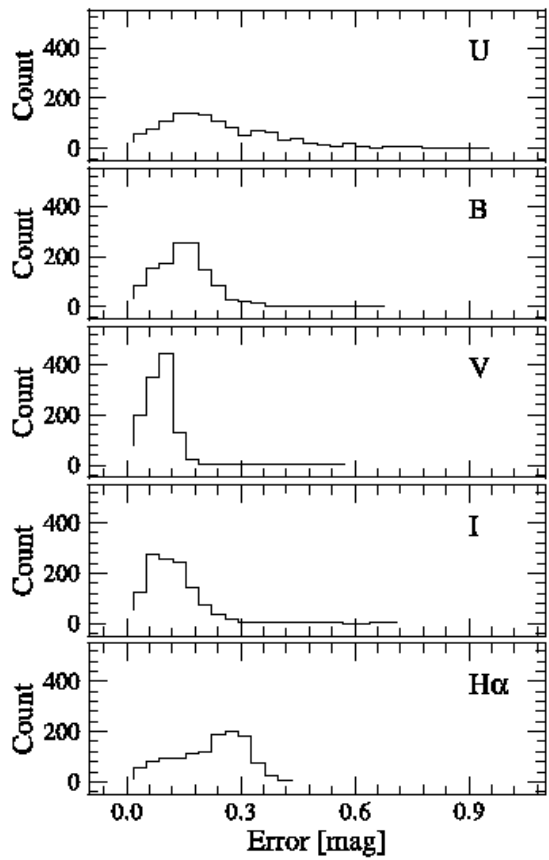}
	\end{center}
	\caption[]
	{Photometric uncertainty distributions for the 1243 M\,83 cluster fluxes in the five
	photometric bands used in this study.
	\label{fig:ObsErrorsDistrib}}
\end{figure}

\citet{Fouesneau2010} assessed the methods used here to estimate cluster ages
and masses. Using standard UBVI or UBVIK photometry and accounting for
observational errors of $5$\% on the fluxes, they showed that it was possible to
determine age distributions with a resolution in age much narrower than in
non-stochastic studies.  In this section, we briefly describe the behaviour of
the analysis when using photometry in the UBVI and \halpha filters of HST/WFC3,
with error bars distributed as in the actual M\,83 cluster data
(Fig.\,\ref{fig:ObsErrorsDistrib}).  Most of these uncertainties are larger than
$5$\%.  This experiment allows us to identify potential artifacts, and to
determine what resolution in age to aim for in the analysis of the empirical
dataset.  Unless otherwise stated, the second method of
\citet{Fouesneau2010} is used in the analysis here and in subsequent sections,
i.e. we maximize posterior probabilities. We refer to
Sect.\,\ref{sec:analysis_extended} for a discussion of a more complete
exploitation of the posterior probability distributions.

The sample of synthetic clusters used as a mock dataset contains 1000 objects
and is built as follows: (i) the number of stars in each cluster is drawn from a
power-law distribution with index $-2$, between $10^3$ and $10^6$ (which
corresponds roughly to masses between $500$ and $5 \times
10^5\,\mathrm{M}_{\odot}$), (ii) the logarithms of the cluster ages are drawn
from a uniform distribution, (iii) the uncertainties are assigned randomly to
uncertainties taken from the list of clusters observed in M\,83 (see
Fig.\,\ref{fig:ObsErrorsDistrib}), and (iv) extinction is added with $A_V$
distributed uniformly between $0$ and $3$. Note that no noise is added to the
synthetic fluxes (the uncertainties assigned are used only in the inversion).
As in \citet{Fouesneau2010}, the physical quantities used in the inversion are
fluxes, not magnitudes or colors.  Errors initially provided in magnitudes are
reinterpreted as symmetric errors on the fluxes. 

\begin{figure}
        \begin{center}
	  \includegraphics[width=\columnwidth]{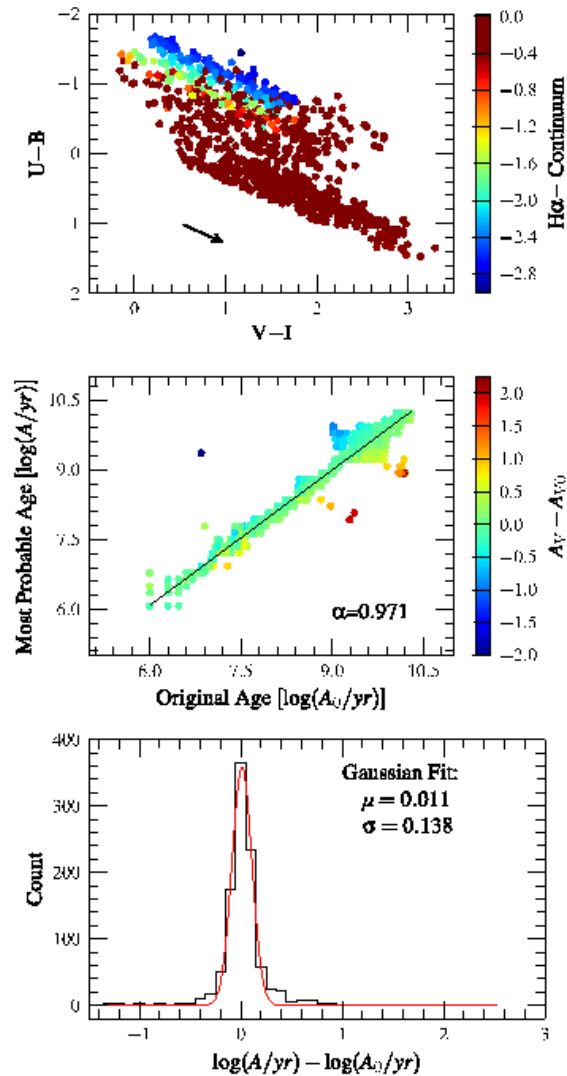}
	\end{center}
	\caption[]
	{Analysis of the colors of the mock data
	(Sect.\,\ref{sec:resolution}).  {\bf Top:} Input sample. Colors
	highlight synthetic clusters with H$\alpha$ line emission.  {\bf Middle
	\& bottom:} Comparison between input ages and derived ages.  The linear
	regression shown has a slope of $0.971$.
	\label{fig:mostProbAgeMasDistribResol}
	}
\end{figure}

The top panel of Figure~\ref{fig:mostProbAgeMasDistribResol} illustrates the
input sample of synthetic clusters. The color scale refers to the
continuum-subtracted \halpha emission of the models\footnote{The continuum
subtraction is performed for this illustration only; total narrow-band
fluxes are used in the analysis.},
which is directly
related to the temperatures and luminosities of the hot bright stars.  This
\halpha emission is significant only for clusters younger than $8\Myr$ (but some
young clusters show none as they happen to contain no ionizing star).  Note that
unlike Fig.~\ref{fig:complete_cmds}, the synthetic clusters shown in
Fig.~\ref{fig:mostProbAgeMasDistribResol} are reddened, hence they trail down
the reddening vector along diagonal lines.  There is a region of overlap in
broad band colors between models with and without \halpha emission (i.e. the
dark red points which represent clusters with no \halpha emission overlap with
the points of other colors which do emit \halpha). Including F657N data is
expected to provide critical information in this regime.

The second and third panels of Fig.\,\ref{fig:mostProbAgeMasDistribResol} show
that ages are recovered without large biases and that errors in the derived
$\log(A/yr)$ have a quasi-normal distribution with a standard deviation of
$~0.14$\,dex  (the global offset of $0.011$\,dex is not significant, it can
be explained by a handful of outliers).   Considering the substructure seen
in the middle panel, we will not attempt to interpret features narrower than
about $0.33$\,dex (FWHM of a Gaussian fit) in the age distributions obtained for
the clusters of M\,83. The equivalent figures for the input and output masses
show a dispersion of about $0.1$\,dex.

Large errors in age occur for a few percent of the synthetic clusters, and
mostly at ages of one or a few Gyr. The origin of these is explained in
\citet{Fouesneau2010} (their Figs. 5, 6 and 8). 
Regions of high model density in color-color space act as attractors in our
analysis.  If the locus of dereddened versions of the photometry of a cluster
approaches such a densely populated region, the predominant age-mass properties
of the models located there will be considered most probable. If the cluster
observed is in fact reddening-free, the result will be an overestimated
extinction together with an underestimated age (the well-known
age-extinction degeneracy is also present in the stochastic context). If the
observed cluster is truly reddened, it sometimes happens that the analysis
underestimates extinction and therefore overestimates age. This last situation,
however, is unlikely to occur above $1\Gyr$ in the M\,83 sample, because most
old reddened clusters are below our detection limit. A small percentage of young
reddened clusters could be affected.

\section{M\,83 cluster ages and masses}

We can now go back to M\,83 and estimate age, mass and extinction for each of
the clusters in the sample in the stochastic context.   As in
\citet{Fouesneau2010}, the assigned estimates are the age-mass-extinction
triplets that maximize posterior probability, i.e.  maximize the probability of
the observed set of fluxes, given the underlying age and mass distributions of
the model collection.  The prior distribution in $\log(age)$ is uniform, the
prior distribution in $\log(mass)$ is also uniform, the metallicity is solar,
and $A_V$ is allowed to take any value between $0$ and $3$ in steps of $0.2$.
We recall that the completeness limits of the observations are not used in any
manner during the determination of individual ages and masses. A discussion
of the errors on the estimates as derived from posterior probability
distributions in age-mass space can be found in
Sect.\,\ref{sec:analysis_extended}.  

\begin{figure}
        \begin{center}
	  \includegraphics[width=\columnwidth]{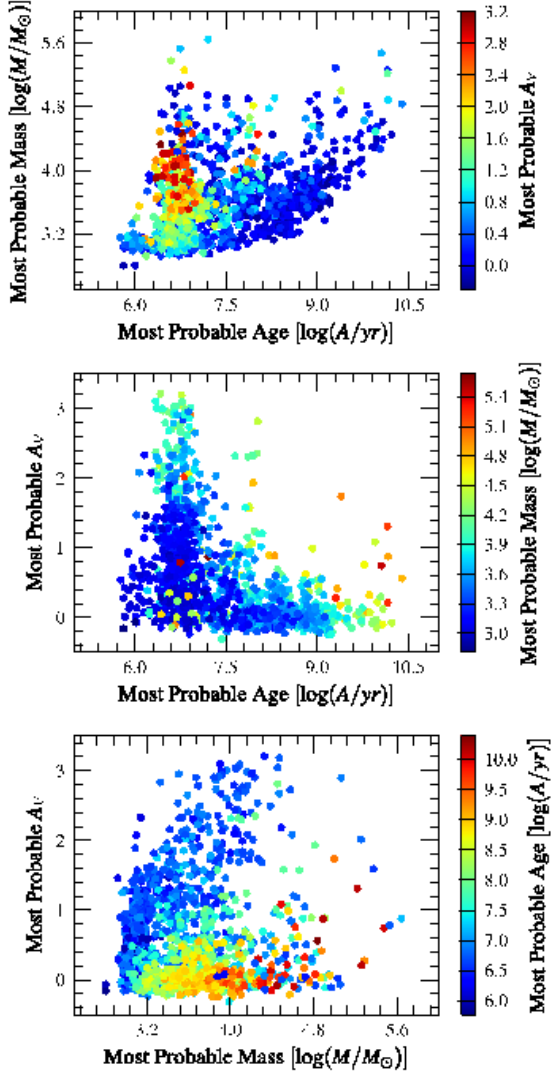}
	\end{center}
	\caption[Bayesian Age--Mass--Extinction distribution]
	{Age--Mass--Extinction distribution obtained with the method
	of \citet{Fouesneau2010}. 
        Noise is added to the dot positions along the different
	directions to reduce the overlap introduced by the
	binning method. A dispersion of $0.15$\,dex in age is used for this
        purpose, which corresponds to the minimal resolution determined in
	Sect.\,\ref{sec:resolution}. A dispersion of $0.05$\,dex is used in
	mass and $0.1$ in A$_V$.
	Colors refer to the third estimated parameter, according to the
	scale on the right-hand side.
	\label{fig:mostProbAgeMasDistrib}
	}
\end{figure}

\begin{figure} 
       \begin{center}
      \includegraphics[width=\columnwidth]{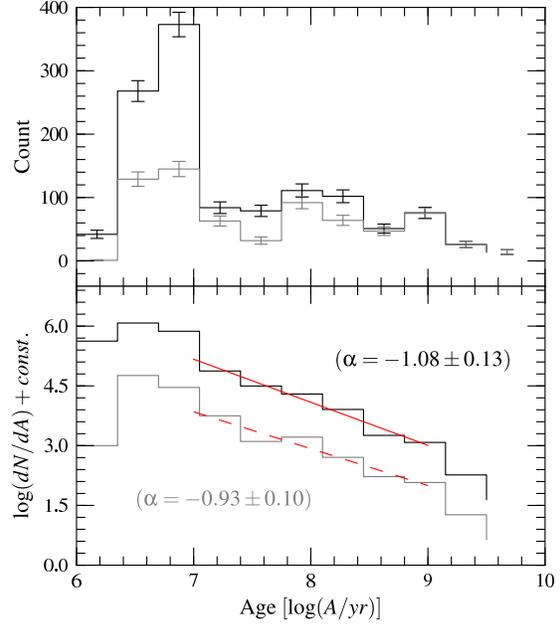}
      \end{center}    
      \caption{Age distribution obtained based on discrete 
      population models. The top panel shows the $dN/d\log(A)$ distribution
      whereas the bottom one shows the $\log(dN/dA)$ distribution, which are
      the two commonly used representations in the literature.
      The black curves represent the distribution for the
      whole set of detected clusters, whereas the gray curves only include
      objects with estimated masses above $10^{3.5}\msun$.
      Error bars are the Poisson noise dispersions in each bin.
      Solid and dashed lines represent power-law fits to the
      distributions of the bottom panels (for $\log(A/yr)$ between $7.0$ and $9.0$). 
      The bin width is of $0.35$\,dex.
      \label{fig:dn_dloga}
      }
\end{figure}

The age-mass-extinction distributions resulting from the analysis are shown in
Fig.~\ref{fig:mostProbAgeMasDistrib}. Derived masses range from the low limit of
our model catalog ($10^3\msun$) to about $5\,10^5\msun$. Ages are distributed
between a few $\Myr$ and about $1\Gyr$, with only a few candidates for ages
older than $1\Gyr$.  The relative lack of old clusters was expected from the
detection limits (Sect.\,\ref{sec:colors_distrib}).  The middle panel of
Fig.~\ref{fig:mostProbAgeMasDistrib} shows that very young clusters come with a
large range of extinction values, as seen in many star forming galaxies
\citep[e.g.][Kim et al.  in prep.]{Whitmore2002}.  At ages older than
$10^7 \yr$, clusters with more than one magnitude of extinction become rare.
Between $10^7$ and $10^8 \yr$, this most likely reflects a real lack of highly
reddened objects, as the data's detection limits in principle allow us to detect
$3\times10^7 \yr$ old clusters with masses above $\log(M) \sim 3.7$ and up to
$2$ magnitudes of extinction. At old ages, we do not expect many highly reddened
clusters to be present in the disk of M\,83. But even if they existed they would
have faded below the detection limits of the sample unless they were very
massive.

The bottom panel is consistent with expectations again: at a given age, the
upper envelope of the derived extinctions decreases with decreasing mass.

\subsection{Cluster ages}
The distribution of all the cluster ages is shown in Figure~\ref{fig:dn_dloga}.
The age distribution of the clusters can be approximated by a power law, with an
index $-0.93 \pm 0.10$ for clusters more massive than $\log(M/\msun)=3.5$ and
with ages between $\approx 10^7-10^9\yr$.  If we fit from $3\times 10^6$ to
$10^9\yr$ instead, we find a power law index of $-1.12 \pm 0.18$.  The slope
remains similar if we restrict the sample to even higher masses. These results
are similar to those found by \citet{Chandar2010}, although they used different
ranges of masses and significantly broader bins to account for age-dating
artifacts resulting from continuous models.

For ages between $10^7$ and $10^9\yr$, the sample's  age distribution is roughly
flat in logarithmic age bins (top panel).  Substructures are observed in some
age distributions with different bin sizes and locations.  These substructures
disappear when the age distribution is plotted with bins wider than $0.4$\,dex.
The distribution drops off at ages older than $10^9\yr$, as expected from the
observational selection limits.

The age distribution of the full sample has a peak at ages of $3-10 \times
10^6\yr$. The excess of clusters found at young ages, compared to a power law
extrapolation of the distribution at older ages, consists of low mass objects
($\log(M/\msun)<3.5$).  Objects {with masses} this small typically have faded
below our detections limits when older than $10^7$\,yr. Imposing a mass cut at
$\log(M/\msun)=3.5$ largely removes this peak.

Systematic errors are likely to affect the strength, width and position of the
peak at young ages.  Indeed, the derived distribution falls to {nearly} zero
below ages of about $3\Myr$, despite the fact that M\,83 is still in the process
of forming stars and clusters.  It is {likely} that some of the clusters with
estimated ages near $5\Myr$ are in fact younger.  Reasons for possible
systematics in this age range include: (i) the sensitivity of the age dating to
the assumed escape fraction of ionizing photons from the region in which the
fluxes are measured {(as discussed in Sect.\ref{sec:colors_distrib})}; (ii) some
level of inconsistency between the stellar evolution models and the clusters in
M\,83; (iii) the fact  that our current model collection contains only models
with ages that are integer multiples of $1\Myr$; (iv)  the lower-mass limit
of the current collection; and (v) our restriction to $A_V < 3$ in the analysis
of cluster colors.  A full usage of the posterior probability distributions
of each individual cluster might reduce some of these systematics (see Sect.
\ref{sec:analysis_extended}).

\begin{figure}
        \begin{center}
	\includegraphics[width=0.8\columnwidth]{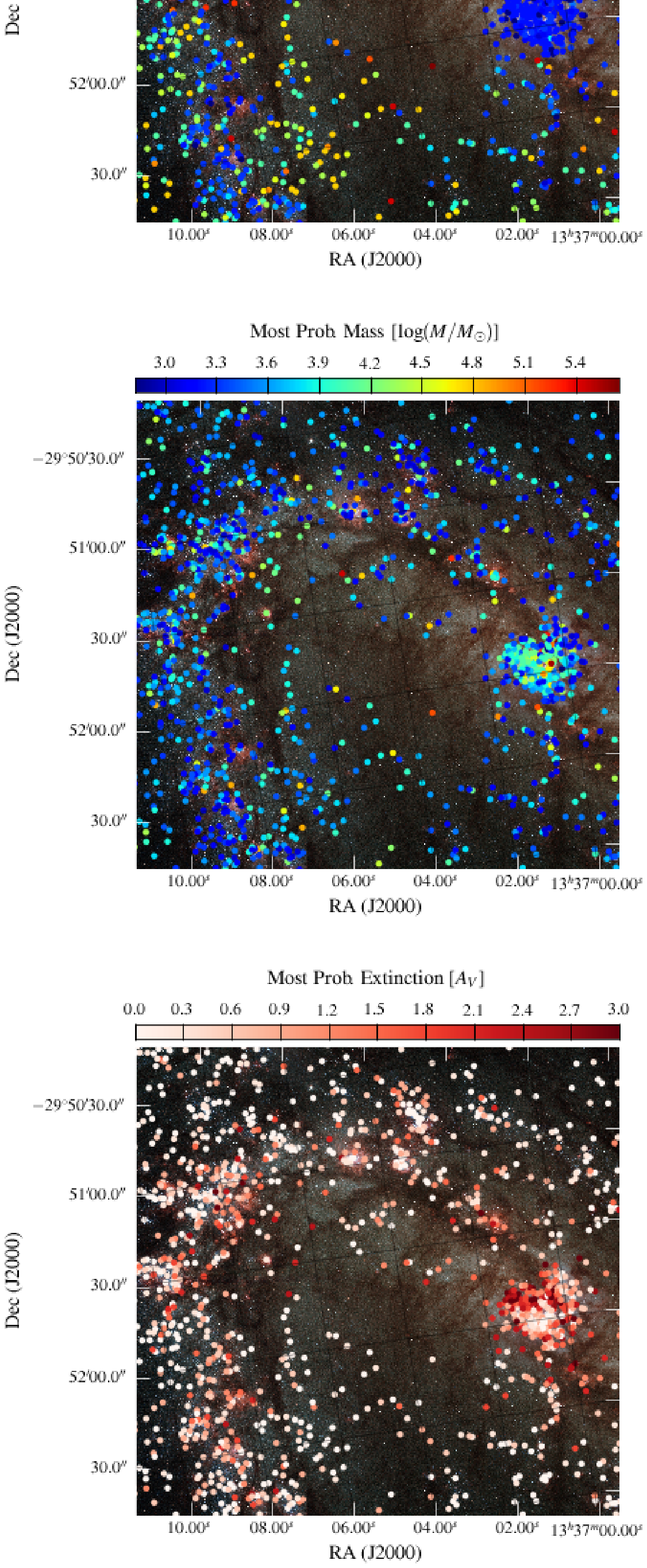}
	\end{center}
	\caption[Spatial distribution of ages, masses, and extinctions]
	{ From top to bottom, recovered ages, masses, and extinctions vs. location
	in the galaxy. Colors are coded according to the respective scale and
	parameter given on the top of each figure.
	\label{fig:spatialDistrib} }
\end{figure}

\begin{figure*}
        \begin{center}
	\includegraphics[clip=]{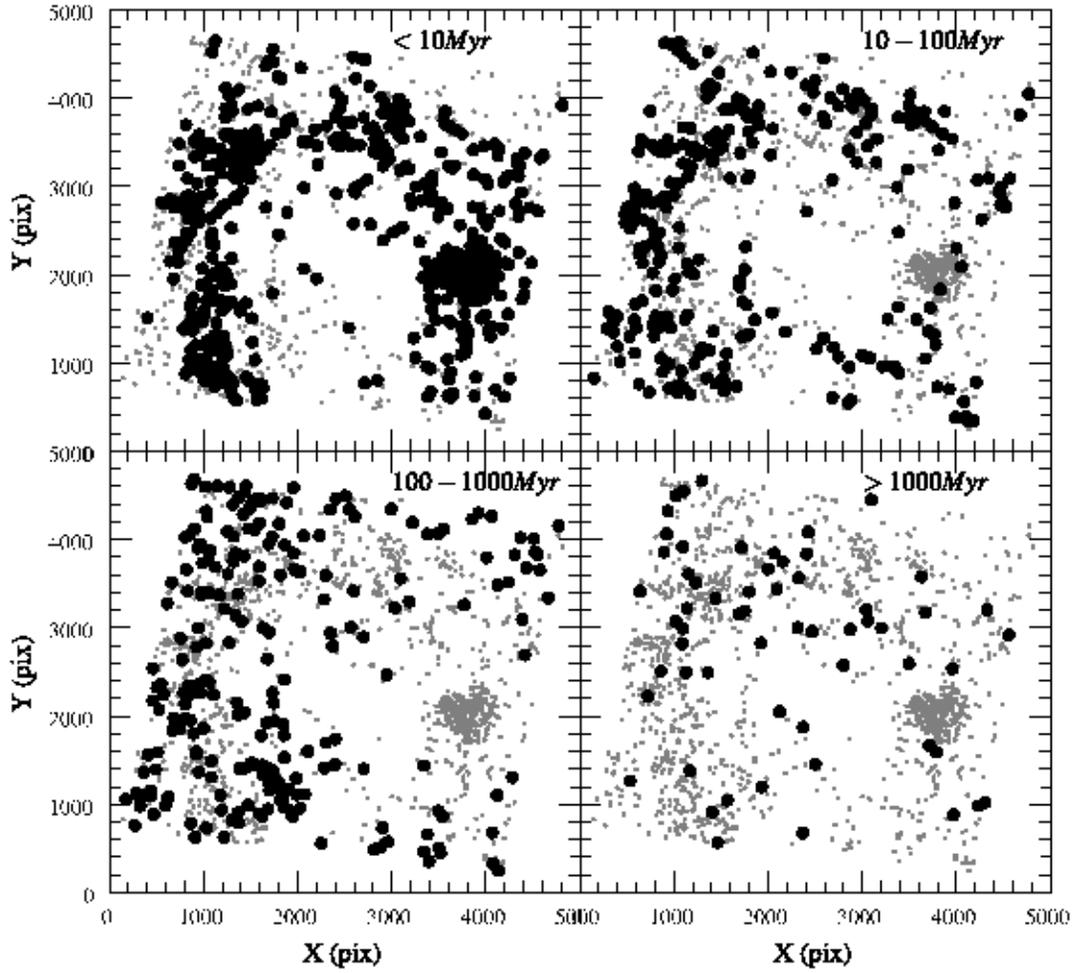}
	\end{center}
	\caption{This figure represents the spatial-age
	distribution of the cluster sample. On each panel, gray dots show the
	whole sample while black dots are objects with ages indicated in the top right
	corner.
	\label{fig:spatialAgeDistrib}
	}
\end{figure*}

The spatial distributions of young and old clusters highlight the effect of
spiral arms (Fig.\,\ref{fig:spatialDistrib}).  As illustrated by
Fig.~\ref{fig:spatialAgeDistrib}, a spatial plot of the subsample of clusters
with assigned ages younger than $10^7\yr$ clearly shows the spiral pattern and
highlights the recent star formation in the nuclear region.  The spiral pattern,
as traced by the clusters, starts fading away for clusters older than
approximately $300\Myr$, and the (only) 64 clusters with ages older than
$10^9\yr$ are consistent with a uniform distribution throughout the disk. 
As selection effects may differ in crowded and reddened areas of the field,
these uncorrected spatial distributions should be taken with caution. 

In summary, the age distribution of clusters younger than $10^9\yr$ and with
masses $\log(M/\msun)$ $>3.5$ is approximately a power law with an index of
$-1.0 \pm 0.2$.  While this is similar to previous results, the distribution is
now defined with better age resolution ($0.4$ vs. $0.7$ dex), and it is based on
more appropriate modelling than previous determinations when masses below
$\log(M/\msun)$ of $4.5$ are included.

\subsection{Cluster masses}

\begin{figure}
        \begin{center}
	\includegraphics[width=\columnwidth]{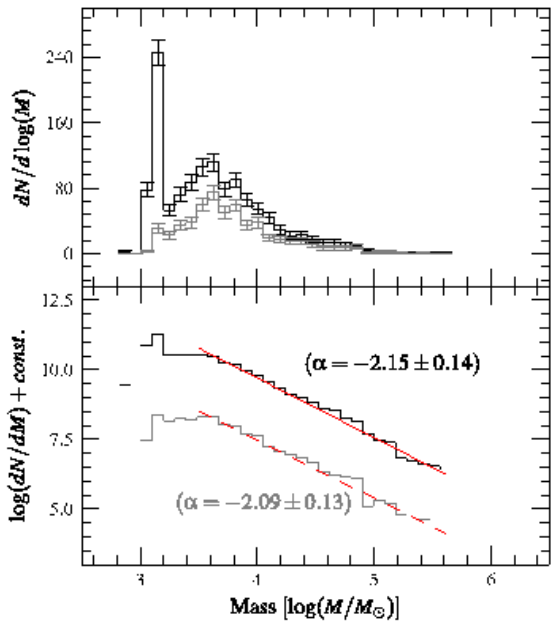}
	\end{center}
	\caption{Mass distribution obtained with the stochastic analysis. This
	is the mass counterpart of the age distributions given in
	Fig.\,\ref{fig:dn_dloga}. The black curves represent the distribution for
	the whole set of identified objects whereas the gray curves only
	includes objects with estimated ages higher than $10\Myr$. Error
	bars are the Poisson noise dispersions in each bin. Power-law fits are
	overlaid on top of the distributions (fitted masses are above
	$10^{3.5}\msun$).   
	\label{fig:dn_dlogm}
	}
\end{figure}

The marginal distribution of all the cluster masses of the sample (i.e. the
distribution summed over all ages and extinctions)  is shown in
Fig.\,\ref{fig:dn_dlogm}.  It can be approximated by a power-law with index
$-2.15 \pm 0.14$ for masses higher than $\log(M/\msun)$ of $3.5$.  The slope
remains within the uncertainties ($-2.09 \pm 0.13$) when we restrict the sample
to clusters with ages between $10^7-10^9$~yr.  This confirms results from
previous analyses, which also found approximatively power-law distributions
\citep{Anders2007, Dowell2008, Fall2009, Fall2005, Chandar2010, Bastian2009}.
Note that the power-law of the prior mass distribution in our analysis is $-1$:
the inversion is able to move away from the prior if necessary.

Below $\log(M/\msun) = 3.5$ the derived mass distribution falls off, and ends
with an apparent peak near $\log(M/\msun) = 3$.  The peak at these low masses is
unreliable, as it corresponds to the lower mass limit of the collection of
synthetic clusters used in the analysis: some of these objects may in fact be
even less massive than $10^3\msun$.  The  relative lack of clusters found
between $\log(M/\msun)=3.2$ and $3.5$ on the other hand is not related to any
known artifact of the analysis
{ (see however Sect.\,\ref{sec:analysis_extended})}.  
It may well result from the incompleteness of
the cluster sample at low fluxes:   based on the Monte-Carlo collection, we
know that log$(M/\msun)=3.7$ is the threshold mass under which more than half
of the  clusters have fluxes below the ``completeness'' limit (defined in
Sect.\,\ref{sec:colors_distrib}) in at least one photometric passband. 

Also, one should recall that 
(i) being faint, the low mass objects have larger observational errors than
massive ones, 
(ii) having low masses while remaining above the detection limits 
these sources  must be young, and therefore we are in the regime most 
sensitive to the modelling of H\,$\alpha$.

In summary, the use of an approach that explicitly accounts for the
stochastic nature of the observed clusters confirms a power-law behaviour for
the mass distribution of the M\,83 clusters, with an index of approximately
$-2.1 \pm 0.2$, over the range of masses least affected by incompleteness, i.e.
between $\log(M/\msun)=3.5$ and $\log(M/\msun)=5$.  

\subsection{Nucleus versus disk}

The $\sim200$ star clusters of our sample that are part of the galaxy nucleus
have derived ages younger than $10\Myr$ with our analysis ($\sim 4\Myr$ on
average).  This statement is not to say the nucleus contains no older objects,
since {most} older clusters would remain undetected against the high background
of the young sources. The higher average mass of the nuclear clusters ($\sim
10^{4}\msun$) is also compatible with the selection effects expected in an area
crowded with bright objects.


Our main conclusions about the age distribution and the mass distribution of the
M\,83 sample remain valid for the disk when the nuclear subsample is removed.
The age distribution is changed only below $10\Myr${, hence the fits shown in
Fig.~\ref{fig:dn_dloga} are essentially unaffected}. The field clusters' mass
distribution is only slightly depleted at high masses compared to the mass
distribution of the whole sample, and this steepens the index of the fitted
power law by only a few percent.

\section{Discussion} \label{sec:discussion}

\subsection{The importance of the \halpha filter, F657N}\label{sec:halpha}

Despite the importance of \halpha in the identification and age-dating of young
clusters, this filter is still not used extensively. This is partly because
extra difficulties come with these measurements and their interpretation. For
example, the spatial extent of HII regions rarely matches that of the underlying
optical continuum light leading to difficult aperture corrections, the escape
fraction of ionizing photons is poorly known, the hot ionizing stars are a rare
subpopulation and their number in any given cluster can deviate strongly from
the expected mean \citep{Cervino2003}. Considering these difficulties, we tested
the results of the analysis of the M\,83 clusters with UBVI data alone, i.e.
discarding the flux measurement in the \halpha filter (F657N). 

\begin{figure}
        \begin{center}
        \includegraphics[width=\columnwidth]{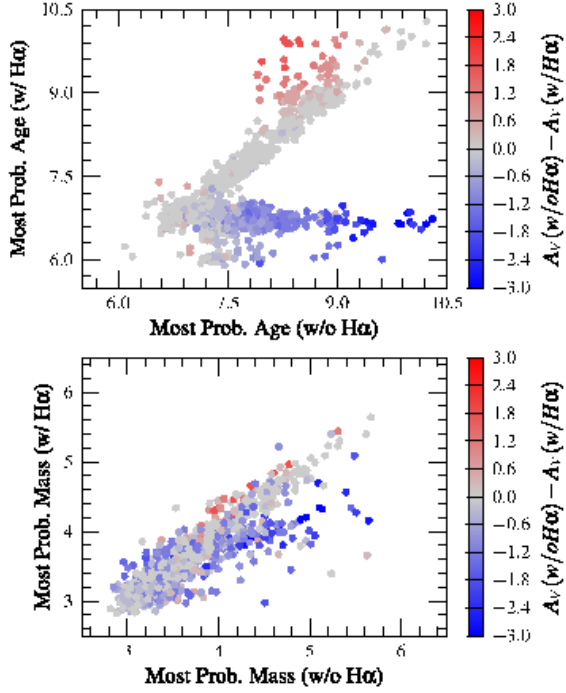}
        \end{center}
        \caption{Comparing age and mass estimates (from the stochastic approach)
        based on UBVI+H\,$\alpha$ (y-axes) and UBVI only (x-axes). Each dot
        represents the estimates of a star cluster color-coded according
        to the change in the extinction estimate (resulting from the inclusion
        of H\,$\alpha$). Age estimates are compared on the top panel, while mass
        estimates are compared on the bottom panel.
        Projected age and mass
        distributions are represented in Fig.\ \ref{fig:halphaCompDist}.
        \label{fig:halphaComp}
        }
\end{figure}

\begin{figure}
        \begin{center}
        \includegraphics[width=\columnwidth]{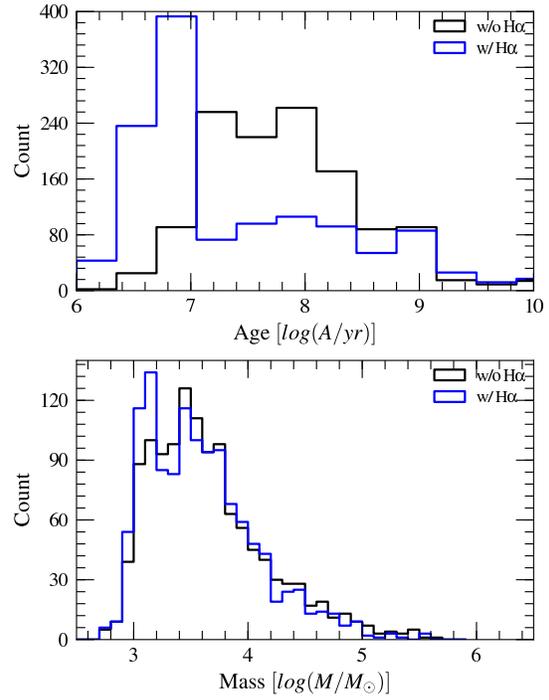}
        \end{center}
        \caption{Comparing age and mass distributions (from the stochastic approach)
        based on UBVI+\halpha (blue) and UBVI only (black).
        Age estimates are compared on the top panel, while mass
        estimates are compared on the bottom one. 
        \label{fig:halphaCompDist}
        }
\end{figure}

Figure~\ref{fig:halphaComp} shows how the ages and masses estimated with
\halpha included ($y$-axis) compare with those obtained from UBVI alone
($x$-axis).  In this figure, the colors represent changes in the estimate of the
extinction parameter A$_V$ between the two studies. The age-extinction
degeneracy is apparent: younger ages come with naturally bluer colors on
average, which can be compensated with a higher A$_V$. The blue dots show that
without the narrow-band filter information many clusters with \halpha
emission are given intermediate ages ($40$\%).  The masses of the corresponding
clusters are systematically lowered when the \halpha filter is included:
the extra extinction required to match the colors at the now younger ages does
not quite compensate for fading (younger clusters are intrinsically brighter for
a given mass).  The red dots also illustrate the importance of R-band
information. Indeed when there is no \halpha emission, the F657N filter becomes
essentially a narrow R-band filter and acts as an extra constraint on the ages
of clusters too old to contain ionizing stars.

Figure~\ref{fig:halphaCompDist} presents the projected age and mass
distributions. Clearly, the age distribution depends strongly on whether or not
\halpha narrow-band (F657N) data is available.  However the effect on the
mass distribution is limited: in particular, we conserve a similar power-law
index.  This suggests that the mass distribution of the sample based on our
probabilistic approach is determined robustly down to about
$\log(\mathrm{M/\msun})=3.5$.

\begin{figure}
        \begin{center}
        \includegraphics[width=\columnwidth]{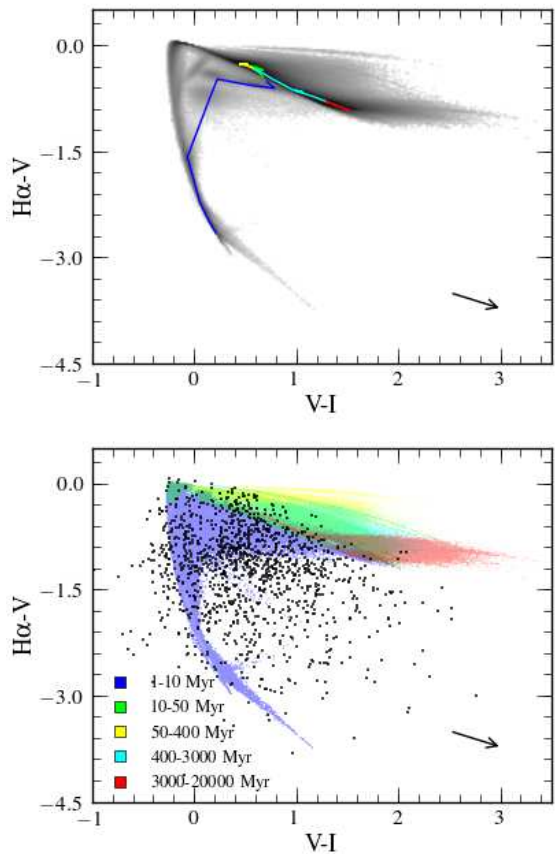}
        \end{center}
        \caption{Models and M\,83 cluster data in a 
        color-color plane that highlights the role of F657N.
	The model distribution is shown as a 2D histogram in the
        top panel, with the line of {\em continuous} models 
        overlaid. The models are shown as a
	colored contour maps in the bottom panel, together
	with the M\,83 data.
        \label{fig:halpha_colcoldiag}
        }
\end{figure}

The age distributions with and without F657N differ, but which one is closer to
reality? How sensitive are the young ages to details in the modelling or errors
in the measurements?  A color-color diagram involving F657N helps answering this
question (Fig. \ref{fig:halpha_colcoldiag}).  The locus of the cloud of data
points is extremely well represented by the collection of stochastic models with
the extra degree of freedom provided by the reddening vector (this is reflected
in the very good best-$\chi^2$ values obtained with the stochastic
models).  { As opposed to the continuous models, the stochastic ones reproduce
the ``corner'' of the observed distribution at (V-I)$\simeq$0 and
(\halpha$-$V)$\simeq 0.$}

Errors would have to be unrealistically large and systematic to avoid an
interpretation of Fig.\,\ref{fig:halpha_colcoldiag} with a large fraction of
objects younger than $10\Myr$.  Changing the escape fraction of ionizing photons
in the models would modify only the extent and color of the ``plume'' of young
models in the figure. Changing the slope of the upper stellar IMF, the exact
ages for which models are computed, or the prior age and mass distributions of
the model collection, would change how models are distributed along this plume
and near the blue ``hook'' in the distribution. The slope of the extinction
vector determines where exactly the dereddening line of a cluster reaches the
cloud of models. But none of these changes would affect the fraction of young
clusters greatly.

{To summarize,}  the conclusion that about $55$\,\% of the clusters in the
sample are less than $10^7$\,yr old is robust. On the other hand, individual
ages below $10^7$\,yr are sensitive to model details.

\subsection{Comparison with the results based on continuous models}
\label{sec:ContVsStoch}

\begin{figure}[ht]
        \begin{center}
        \includegraphics[width=\columnwidth]{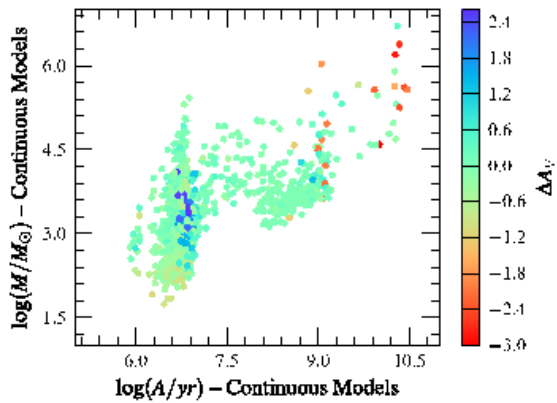}
        \end{center}
        \caption[Age--Mass distribution obtained with continuous models]
        {Age-mass distribution obtained from a $\chi^2$-fit using
        continuous models.  Colors other than cyan and green
        highlight clusters for which the ages have changed most strongly
        between the continuous and the stochastic analysis (actually
        the scale is set by the corresponding change in A$_V$).
	\label{fig:StdAgeMasDistrib}}
\end{figure}

For a direct comparison, we have repeated the analysis of the M\,83 clusters
with the traditional approach,  based on the mean fluxes predicted by
continuous models.  The population synthesis assumptions are those of {\sc
P\'egase} as described above, and the $\chi^2$ calculation used to  measure the
quality of a fit is defined using fluxes as in the stochastic context
\citep[see][who present such a comparison for synthetic
clusters]{Fouesneau2010}.

The 2D age-mass distribution resulting from the ``continuous'' analysis is shown
in Fig.\,\ref{fig:StdAgeMasDistrib}. When compared to the results of the
stochastic analysis (top panel of Fig.\,\ref{fig:mostProbAgeMasDistrib}) it
shows a less homogeneous distribution in age. \citet{Fouesneau2010} explained
why this is expected. 

\begin{figure*}[ht]
\begin{center}
\includegraphics[width=2\columnwidth]{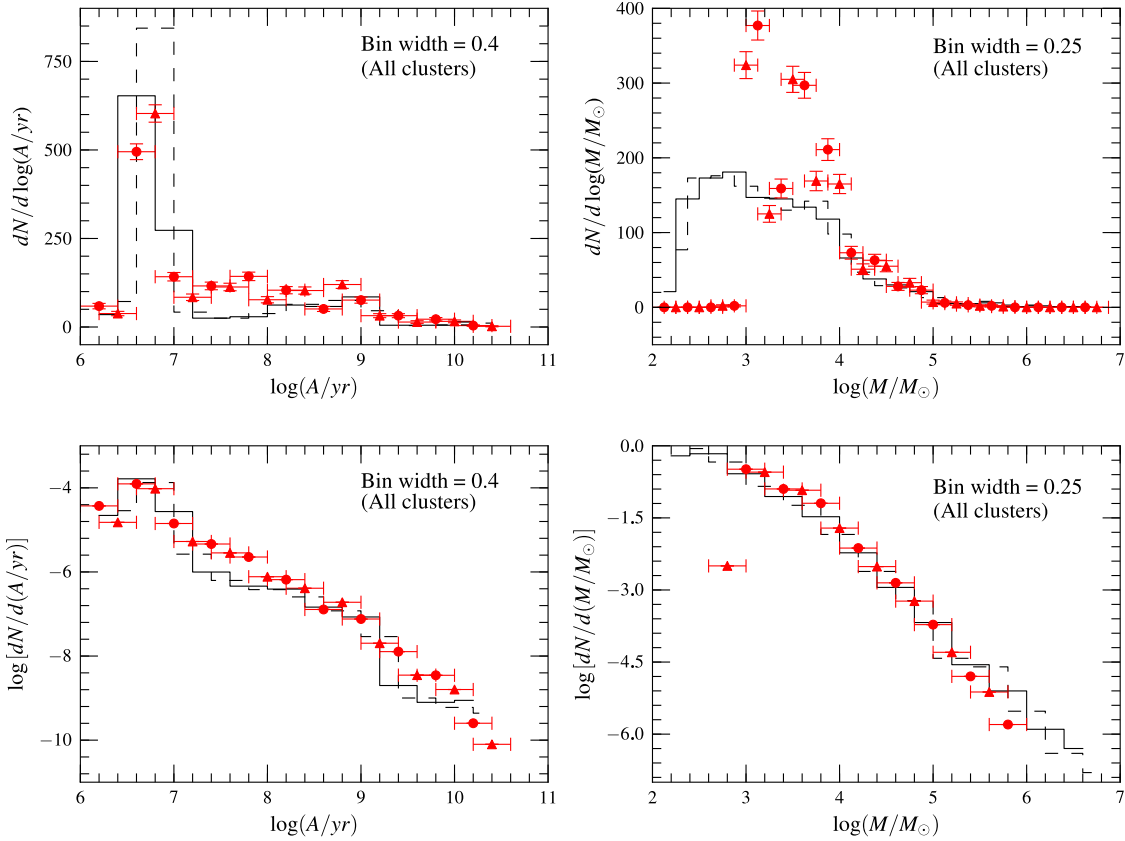}
\end{center}
\caption[Marginal distributions, continuous and stochastic models]{
 Marginal distributions of cluster ages and masses as obtained  with
 continuous models (solid and dashed histograms) and with stochastic
 models (red histogram levels). In each case, the histograms are
 shown for two sets of bin centers, offset from each other by half a
 bin size. To guide the eye, error bars representing Poisson standard deviations
are shown in the top panels.
\label{fig:MarginDistribContVsStoch}}
\end{figure*}

The ``continuous'' marginal distributions of cluster masses and cluster ages are
compared to the stochastic ones in Fig.\,\ref{fig:MarginDistribContVsStoch}.  As
already stated in the previous section, we consider the flatness of the
``stochastic'' age distribution between $10^7$ and $10^9$\,yr {(in logarithmic
bins)} a robust result, while the dip in the ``continuous'' age distribution
around $5\,10^7$\,yr and the excessive accumulation of very young clusters are
artifacts of that method. 

To first order, the mass distributions in the bottom right panel of
Fig.\,\ref{fig:MarginDistribContVsStoch} are alike.  Above $\log(M/\msun)=4$,
they are within 2\,$\sigma$ of each other ($\sigma$ representing Poisson
standard deviations in the bins).

A closer look shows that the analysis with continuous models leads to a subset
of very young and low mass clusters ($\log(M/\msun) < 3$) that are not found
with the ``stochastic'' analysis. At first glance, one may think this
difference simply reflects a bias of the ``stochastic'' analysis, due to the
lower mass limit of the model collection ($10^3\msun$).  In fact, it is due at
least in part to a known issue of the analysis of finite clusters with
continuous models \citep[][their Fig.\,9]{Fouesneau2010}.   A purposely
designed test was run with the WFC3 filters of the M\,83 observations. The
analysis of synthetic clusters, all more massive than $10^3\msun$ and brighter
than the magnitude limits of the M\,83 observations, shows that $\sim$35\,\% of
the clusters with actual $\log(M/\msun)$ between $3.2$ and $3.7$ are assigned
masses below $\log(M/\msun)=3$ when analysed with continuous models.  In the
upper right panel of Fig.\,\ref{fig:MarginDistribContVsStoch} we see that the
difference between the distributions derived with stochastic and continuous
models for  $3.2 < \log(M/\msun) < 3.7$ represents just about $\sim35$\,\% of
all the clusters in that mass range.

More details on the differences between the stochastic and the continuous
analysis are given in Appendix A.  In particular, the one-by-one comparison of
the ``continuous'' and the ``stochastic'' ages and masses are shown. Individual
ages differ by more than $0.3$\,dex for 28\,\% of the sample. 

A comparison between the ``continuous'' results found {in the present study
(i.e. based on {\sc P\'egase} models analog to the stochastic models)} and those
of \citet{Chandar2010} is provided in Appendix B. Using different population
synthesis models changes the ages of 25\,\% of the clusters by more than
0.3\,dex, but in a way that does not modify global results.

\subsection{Comparison with morphological or spectroscopic ages}
\label{sec:Spectro}

\begin{figure}

        \begin{center}
                \includegraphics[width=0.9\columnwidth]{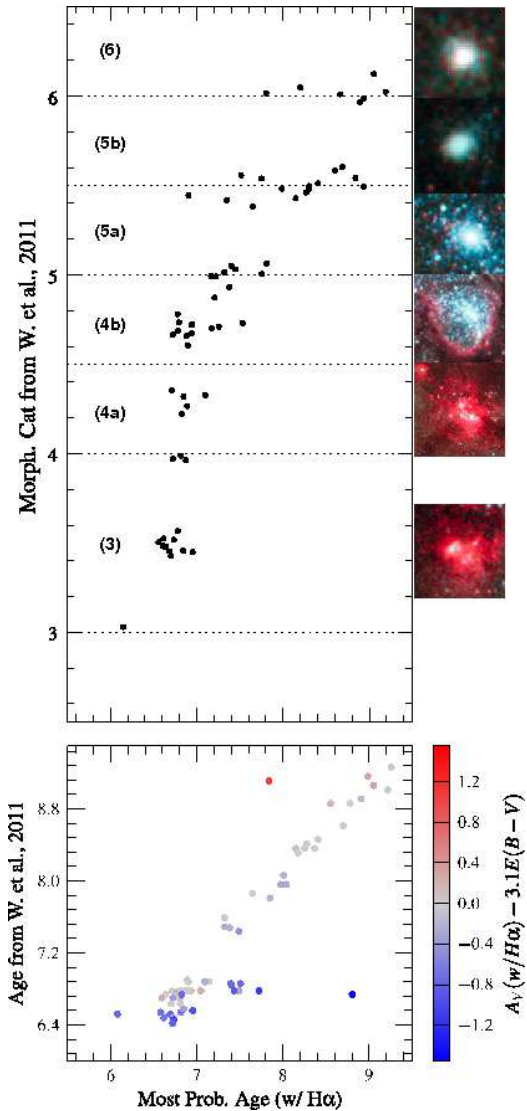}
        \end{center}
	\caption[Whitmore cats vs. this paper]{ Ages derived from the 
	approach detailed in this paper [$x$-axis] compared to the \halpha morphology {categories} from
	\citet{Whitmore2011} (top) and to ages from
	traditional models (bottom), for the clusters in common. Boxes on the
	top panel refer to statistics on the morphological categories presented
	in the cited paper.
	\label{fig:WhitmoreCat1}}
\end{figure}

Age estimates based on methods other than the analysis of the broad and
narrow-band Spectral Energy Distributions (SED) are available for small subsets
of the star clusters of our sample. Unfortunately these subsets do not
extend to the lowest cluster masses. 

\citet{Whitmore2011} have devised a purely morphological classification scheme
for star clusters based on images in several photometric pass-bands, including
H$\alpha$.  The categories are chosen such as to represent subsequent phases of
the morphological evolution of a typical cluster, and the classification
sequence is therefore expected to correlate with age. Their Fig.\,$3$ shows that
the morphological category indeed relates to the SED ages estimated using
traditional models with a continuously populated IMF. 

Fig.\,\ref{fig:WhitmoreCat1} focuses on the $64$ star clusters of
\citet{Whitmore2011} for which the morphological classification is considered
most robust.  Eight of the $64$ clusters are located in the nucleus, $56$ in the
field.  The bottom panel compares our age estimates with the photometric ages
obtained by \citet{Whitmore2011} based on continuous models. Both of these age
estimates use UBVI+\halpha photometry.  The two sets of photometric ages agree
well, except for a scarcely populated horizontal branch of ``catastrophic
differences'' {around $\log(A/yr)=6.9$ from \citet{Whitmore2011} } (only
$10$\,\% of the sample) and an associated gap in the one-to-one relation {around
$\log(A/yr)=7.2$,} a behaviour rather typical for this sort of plot
\citep{Fouesneau2010}.  The deviant clusters are in the morphological categories
$5a$ and $5b$ according to \citet{Whitmore2011}, i.e. they are unlikely to be as
young as the photometric analysis with continuous models suggests.

The top panel of Fig.\,\ref{fig:WhitmoreCat1} shows the relationship between
photometric age and morphological category obtained with the new photometric age
estimates, i.e. those based on stochastic models.  The clusters that are
outliers in the upper panel were also outliers in the version of this figure
given by \citet{Whitmore2011}: they formed a vertical extension of young
photometric ages into older morphological categories.  But for one object, this
vertical extension has now disappeared.  As a result, the average photometric
ages of categories $5a$  and $5b$ shift to older values by about $0.1$\,dex.
This remains within the error bars associated with the dispersion within the
categories.  All in all, the figure remains similar to its original version.

\medskip
Spectroscopic age estimates are available for 13 luminous clusters in the
nuclear region from \citet{Wofford2011}. These were obtained using STIS FUV
spectra covering the range from $1200-1700$ \AA\ (e.g., including the strong
\ion{N}{5} and \ion{C}{4} lines as well as several other weaker lines).
\citet{Wofford2011} fit the observations using both semi-empirical models, based
on a library of Galactic O and B stars observed with IUE, and theoretical
Starburst 99 models \citep[continuously sampled IMF]{Leitherer2009}. They find
ages between $2$ and $5\Myr$ for nine clusters and ages between $10$ and $20
\Myr$ for four.  The photometric ages quoted in that article, based on the same
five photometric pass-bands as in our paper and on the method of
\citet{Chandar2010}, are in good agreement with the spectroscopic ages except
for one of the four objects with spectroscopic ages above $10 \Myr$
\citep[see][Table $3$]{Wofford2011}.  Our analysis with stochastic models
returns ages between $3$ and $9 \Myr$ for all the $13$ clusters. These ages are
within $0.2$\,dex of the spectroscopic results except for the three oldest
objects of \citet{Wofford2011}. The differences for these three objects are due
to differing assumptions in the population synthesis codes rather than to
stochastic modelling.  Indeed, when we use models with a continuously sampled
IMF we obtain ages within $0.1$\,dex of those derived in the stochastic context,
as long as both are based on the same synthesis parameters (e.g. those defined
in Sect.~\ref{sec:models}). This comparison is meaningful because the three
clusters are massive (masses in the range $5\times 10^4 - 1\times 10^6\msun$
depending on the method used).  All in all, the agreement between stochastic and
non-stochastic ages is very satisfactory considering the diversity of
approaches.

\subsection{Towards a more complete usage of posterior probability 
distributions}
\label{sec:analysis_extended}

The main aim of this paper is the comparison of results of traditional
analysis methods with results based on the method 
of \citet{Fouesneau2010}, for a sample of real clusters in M\,83. 
{ Following these authors, we have 
chosen the peak of the posterior probability distribution 
to assign an age, a mass and an extinction value to each cluster. 
This method can be extended to use the full posterior probability 
distribution for each cluster rather than the peak approximation.
In this section, we explore this approach by comparing 
the results of this {\em extended} method with those of the {\em peak} method 
for our cluster sample in M\,83. 
A complete theoretical study of the extended method
in various observational regimes, will deserve a separate paper.}

{ The 3-dimensional probability distributions of age, mass and extinction 
obtained for the 1242 individual clusters  of our sample are quite complex 
and display a variety of patterns. We have found about a dozen typical 
behaviours, that depend on the region of color-luminosity space the clusters 
are in, as well as on observational error bars. 
It is difficult to find a satisfactory way of translating this
complexity into simple error bars on the estimates provided 
by the ``peak" method.
As already described in \citet{Fouesneau2010}, the main trends are set by the
age-mass and age-extinction degeneracies, which define directions along which
probability peaks tend to be elongated.  Within this gross picture
other patterns are seen.  For instance, it is relatively common 
to find bimodal distributions with one peak at young ages ($<10^7$\,yr) and 
one at intermediate ages($\sim 10^9$\,yr). The relative strengths
of these peaks depend on the error bars on the H$\alpha$ measurement. 
This particular pattern can be understood with the discussion we 
provide in Appendix A concerning the left panel of 
Fig.\,\ref{fig:CvsStoch_interpret}.  
}

\begin{figure}
\includegraphics[width=\columnwidth]{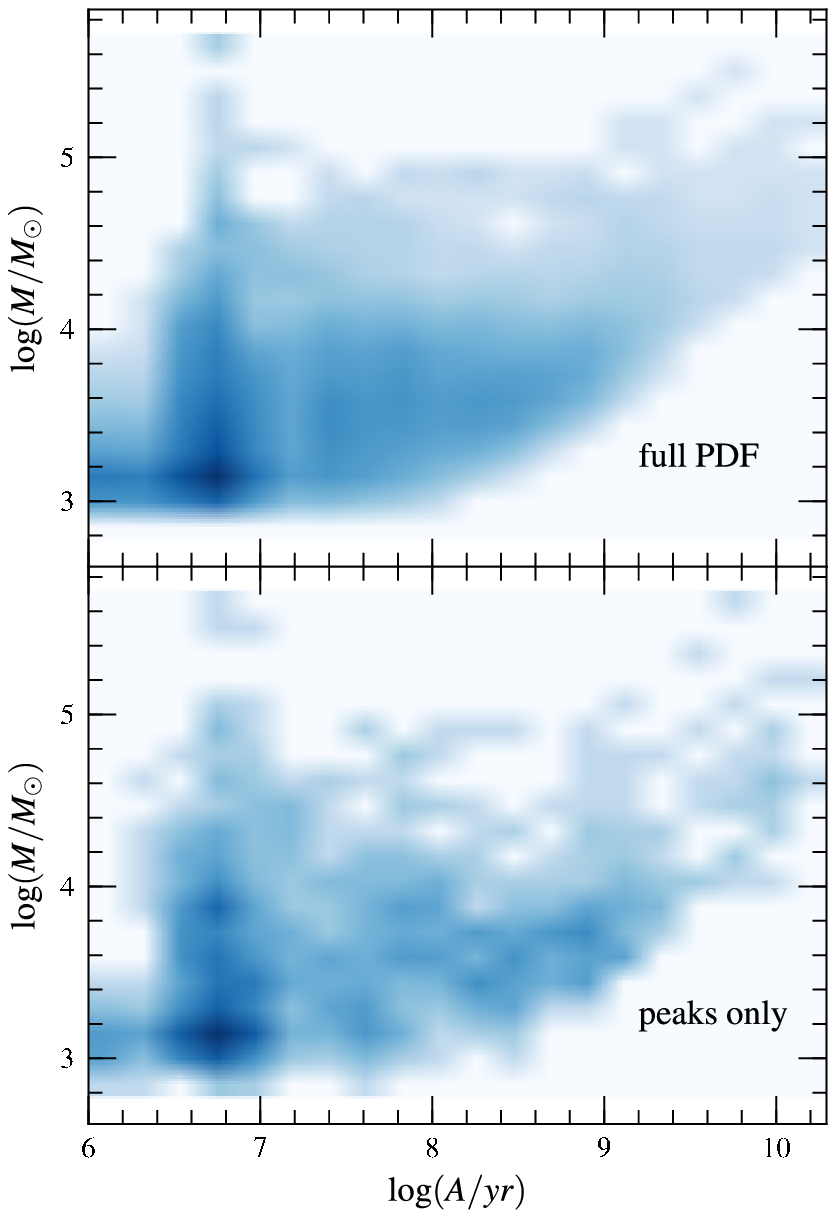}
\caption[]{
Age-mass distributions obtained for the M\,83 sample from the probabilistic
analysis.  {\bf Top}: distribution obtained by summing the individual posterior
probability maps of the clusters (i.e. the {\em extended} method).  {\bf
Bottom}: density map obtained from the most probable age and mass values shown
in the top panel of Fig. 5 (i.e the {\em peak} method, for which each point is
replaced by a patch to produce a density representation).}
\label{fig:AgeMass_extended}
\end{figure}

A new estimate of the age and mass distributions of the cluster sample 
as a whole was obtained by summing the age-mass probability distribution 
of individual clusters (marginalized over extinction). 
{ With this approach, uncertainties on the estimated 
parameters and any covariance between them are automatically
accounted for. The sensitivity to binning choices is reduced.} 
The resulting map is shown in
Fig.\,\ref{fig:AgeMass_extended}.  For convenience, the figure also recalls the
results of the ``peak" method (Fig.\,\ref{fig:mostProbAgeMasDistrib}) as a
directly comparable density map.  The global structure with both approaches is
similar.  For instance, both methods obtain an over-density of young, low-mass
objects. Overall, by accounting for the entire posterior probability
distributions instead of only the location of their maxima we smear out local
variations in the derived age-mass map.

\begin{figure*}
\begin{center}
\includegraphics[width=2\columnwidth]{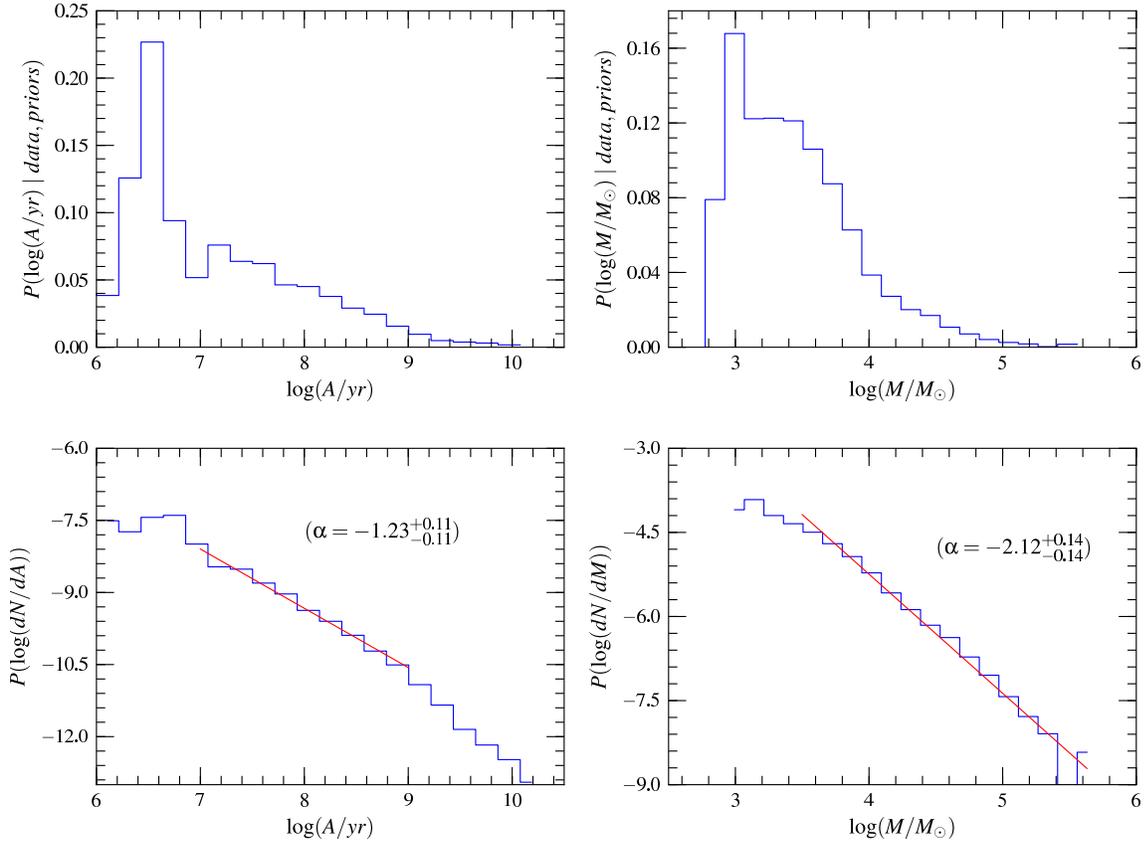}
\end{center}
\caption[]{ 
 Age and mass distributions obtained with the approach explored in
Sect.\,\ref{sec:analysis_extended}.  Left: age distribution, for comparison with
Fig.\,\ref{fig:dn_dloga}.  Right: mass distribution, for comparison with
Fig.\,\ref{fig:dn_dlogm}.  }
\label{fig:OneDdistribs_extended}
\end{figure*}

The one-dimensional age and mass distributions obtained with the extended
approach are presented in Fig.\,\ref{fig:OneDdistribs_extended}, and can be
compared to those shown in Figs.\,\ref{fig:dn_dloga} and \ref{fig:dn_dlogm}. 
The masses of the cluster sample are found to follow a power-law with an 
index of $-2.1$ for $\log(M/\msun)>3.5$, 
a result essentially identical to the one we obtained 
with the peak method.  The restriction to ages older than 10\,Myr 
does not change the index value significantly. 
{ The dip seen near $\log(M/\mathrm{M}_{\odot})=3.3$ with the 
peak method has been smoothed into a more robust plateau.} 
The age distribution in
Fig.\,\ref{fig:OneDdistribs_extended} has a somewhat steeper power-law ($\alpha
= -1.23$) than found previously ($-1.08$).  Its restriction to masses greater
than log($M$/M$_{\odot})=3.5$ reduces the absolute value of the age power-law
index by some 10\,\%.  The slopes of the age distributions, as derived
with either the extended and the peak method, 
remain within the formal 1-$\sigma$
confidence interval of each other for this sample.

{ In summary, the age-mass distributions found 
for our sample of M\,83 clusters
are similar whether we use the extended method or the peak method.
Spurious small scale features in the result of the peak method tend
to disappear with the extended method. The extended method deserves
further investigation in the future.}

\section{Conclusions} 

Most star clusters in nearby galaxies have masses below $10^5\msun$, and hence
their integrated fluxes and colors can be strongly affected by stochastic
fluctuations in the number of massive stars (strictly post main-sequence
stars).  \citet{Fouesneau2010} previously introduced new stochastically sampled
stellar population model predictions, and developed a probabilistic method for
estimating the ages, extinctions, and masses of star clusters from these models.
The models were tested using synthetic clusters.  For the purpose of the present
paper we enhanced the stochastic models by including predictions for narrow-band
filters and extending the Monte-Carlo collection of synthetic clusters to
include higher cluster masses.  We compared the predictions from the
stochastically sampled models with observations of a sample of real star
clusters. Our observations consist of integrated UBVI\halpha fluxes of star
clusters in the nearby spiral galaxy M\,83, which was observed with the Wide
Field Camera 3 (WFC3) on the Hubble Space Telescope.  This is the same catalog
of clusters analysed by \citet{Chandar2010} using predictions from more
traditional models which assume a ``continuously sampled'' stellar IMF.

The locus of the collection of stochastic models in color space (e.g., Figure
\ref{fig:complete_cmds}) shows excellent agreement with that of the collection
of cluster observations. Clusters with broad band colors either bluer or redder
than those of the traditional models find a natural match in the stochastic
model collection, in which some synthetic clusters have underpopulated or
overpopulated red post-main sequence branches.  Similarly, the locus of
stochastic models in color-color planes that include \halpha fluxes (e.g.,
Figure \ref{fig:halpha_colcoldiag}), and hence take into account the random
character of the number of ionizing stars in a cluster, provide a natural match
with the observations.

A key result of the study is the importance of including a narrow band filter
measurement in the analysis, which to date has been used in relatively few
works.  A comparison of age and mass estimates with and without the \halpha
filter shows that ~$30$\% of the sources have age estimates that change by more
than $0.3$\,dex in age in our analysis (factor of 2). The errors are systematic:
without the narrow band measurement, many young clusters are not recognized as
such.

The use of stochastic models has allowed us to derive the age-mass distribution
of the star clusters in the M\,83 sample with a better resolution in time than
was possible before, and we extend both {age and mass} distributions to lower
masses (down to $\log(M/\msun) \sim 3.3$). Above $\log(M/\msun)=3.7$, the mass
limit at which completeness issues in the sample become severe, the mass
distribution derived from the analysis with stochastic models closely resembles
the distribution obtained with traditional models in previous work.  More
specifically, the overall slope of the mass distribution remains within the
range of values indicated by \citet{Chandar2010}, i.e. power-law of index near
$-2$. The age distribution is now free of the dip between $10^7$ and $5 \times
10^7$ years that was a known artifact of the analysis of clusters subject to
stochastic fluctuations with continuous models. Overall, we confirm that it
declines more-or-less continuously starting at young ages as already suggested
by the results of \citet{Chandar2010}. Between $10^7$ and $10^9$ years, the data
are consistent with a flat distributions in logarithmic age units{, which
corresponds to $dN/dA \propto A^{-1}$}. The sample contains only a handful of
clusters older than $1 \Gyr$ because these are too faint for easy detection. 

In this paper we also begin the process of extending the analysis from a
unique ``most probable'' value (corresponding to the peak in the posterior
distribution, the {\em peak} approach) to a full probabilistic description of
the correlated age, mass, and extinction properties of each individual clusters
(i.e. the {\em extended} method). { The extended method is expected to
be more robust than the peak method. It does not loose any of the 
information carried by individual posterior probability distributions
about uncertainties, about correlations between parameters, 
or about multiple (almost) equally probable solutions.  
The age-mass distributions obtained with the
extended method for our M\,83 cluster sample remain similar to those obtained
previously. Spurious substructure seen in {\em peak} method distributions
when plotted with narrow bins disappears.
We plan to test the extended method more systematically in 
various observational regimes in the future.}

{ It may seem surprising at first glance that low resolution age
and mass distributions of cluster samples 
are not modified more radically by the move from
a traditional analysis to a method based on explicitly stochastic models.}
Changing a power law slope radically, with data that extends over about
two dex in age and mass, requires strong and systematic changes, but the changes
induced by the switch from traditional models to stochastic models are not of
this nature. \citet{Fouesneau2010} had predicted that overall mass distributions
would be rather insensitive to this switch, and we have confirmed this with the
M\,83 sample. Their figures also indicated that changes in the age assignments
due to the switch to stochastic models would mostly affect subsets of objects in
particular ranges of age and mass (i.e. the high resolution features of the age
distribution), instead of changing the global balance between young,
intermediate and old objects. This is also what we found for M\,83 sample in the
present paper: the $-1$ ($\pm 0.2$) slope of the best fit power law to the age
distribution is a robust result for the sample.

As we push to lower masses and larger samples of fainter clusters, the improved
accuracy and time resolution achievable with the new stochastic methods allows 
us to address new questions, such as local variations among cluster populations
within individual galaxies.  For instance, we find that the subsample of young
clusters in our data set clearly displays the spiral structure of M\,83, while
this structure progressively disappears in samples older than a few hundred
millions of years. The cluster age-mass distribution in the field we have
studied is thus (not surprisingly) partly determined by the location of this
field with respect to the spiral arms. This level of details can now be taken
into account when comparing galaxies or areas within galaxies.

\begin{acknowledgements}
This paper is based on observations taken with the NASA/ESA Hubble Space
Telescope obtained at the Space Telescope Science Institute, which is operated
by AURA, Inc., under NASA contract NAS5-26555. It uses Early Release Science
observations made by WFC3 Science Oversight Committee.
The analysis of the results made extensive usage of the Topcat
software, available under General Public Licence from\\
\url{http://www.starlink.ac.uk/topcat/}.\\
{ We thank the anonymous referee for their careful reading and suggestions
which improved our manuscript.}
\end{acknowledgements}

\bibliographystyle{apj}
\bibliography{m83}

\begin{thebibliography}{45}
\expandafter\ifx\csname natexlab\endcsname\relax\def\natexlab#1{#1}\fi

\bibitem[{{Anders} {et~al.}(2007){Anders}, {Bissantz}, {Boysen}, {de Grijs}, \&
  {Fritze-v.~Alvensleben}}]{Anders2007}
{Anders}, P., {Bissantz}, N., {Boysen}, L., {de Grijs}, R., \&
  {Fritze-v.~Alvensleben}, U. 2007, \mnras, 377, 91

\bibitem[{{Barbaro} \& {Bertelli}(1977)}]{Barbaro1977}
{Barbaro}, C., \& {Bertelli}, C. 1977, \aap, 54, 243

\bibitem[{{Bastian} \& {de Mink}(2009)}]{Bastian2009}
{Bastian}, N., \& {de Mink}, S.~E. 2009, \mnras, 398, L11

\bibitem[{{Bastian} {et~al.}(2011){Bastian}, {Adamo}, {Gieles}, {Lamers},
  {Larsen}, {Silva-Villa}, {Smith}, {Kotulla}, {Konstantopoulos}, {Trancho}, \&
  {Zackrisson}}]{Bastian2011}
{Bastian}, N., {et~al.} 2011, \mnras, 417, L6

\bibitem[{{Billett} {et~al.}(2002){Billett}, {Hunter}, \&
  {Elmegreen}}]{Billett2002}
{Billett}, O.~H., {Hunter}, D.~A., \& {Elmegreen}, B.~G. 2002, \aj, 123, 1454

\bibitem[{{Bohlin}(2007)}]{Bohlin2007}
{Bohlin}, R.~C. 2007, in Astronomical Society of the Pacific Conference Series,
  Vol. 364, The Future of Photometric, Spectrophotometric and Polarimetric
  Standardization, ed. {C.~Sterken}, 315

\bibitem[{{Bressan} {et~al.}(1993){Bressan}, {Fagotto}, {Bertelli}, \&
  {Chiosi}}]{Bressan1993}
{Bressan}, A., {Fagotto}, F., {Bertelli}, G., \& {Chiosi}, C. 1993, \aaps, 100,
  647

\bibitem[{{Bruzual}(2002)}]{Bruzual2002}
{Bruzual}, G. 2002, in IAU Symposium, Vol. 207, Extragalactic Star Clusters,
  ed. {D.~P.~Geisler, E.~K.~Grebel, \& D.~Minniti}, 616

\bibitem[{{Cardelli} {et~al.}(1989){Cardelli}, {Clayton}, \&
  {Mathis}}]{Cardelli1989}
{Cardelli}, J.~A., {Clayton}, G.~C., \& {Mathis}, J.~S. 1989, \apj, 345, 245

\bibitem[{{Cervi{\~n}o} \& {Luridiana}(2006)}]{Cervino2006}
{Cervi{\~n}o}, M., \& {Luridiana}, V. 2006, \aap, 451, 475

\bibitem[{{Cervi{\~n}o} {et~al.}(2003){Cervi{\~n}o}, {Luridiana}, {P{\'e}rez},
  {V{\'{\i}}lchez}, \& {Valls-Gabaud}}]{Cervino2003}
{Cervi{\~n}o}, M., {Luridiana}, V., {P{\'e}rez}, E., {V{\'{\i}}lchez}, J.~M.,
  \& {Valls-Gabaud}, D. 2003, \aap, 407, 177

\bibitem[{{Chandar} {et~al.}(2010){Chandar}, {Whitmore}, {Kim}, {Kaleida},
  {Mutchler}, {Calzetti}, {Saha}, {O'Connell}, {Balick}, {Bond}, {Carollo},
  {Disney}, {Dopita}, {Frogel}, {Hall}, {Holtzman}, {Kimble}, {McCarthy},
  {Paresce}, {Silk}, {Trauger}, {Walker}, {Windhorst}, \&
  {Young}}]{Chandar2010}
{Chandar}, R., {et~al.} 2010, \apj, 719, 966

\bibitem[{{Converse} \& {Stahler}(2011)}]{Converse2011}
{Converse}, J.~M., \& {Stahler}, S.~W. 2011, \mnras, 410, 2787

\bibitem[{{de Grijs}(2009)}]{deGrijs2009}
{de Grijs}, R. 2009, \apss, 324, 283

\bibitem[{{Deveikis} {et~al.}(2008){Deveikis}, {Narbutis}, {Stonkut{\.e}},
  {Brid{\v z}ius}, \& {Vansevi{\v c}ius}}]{Deveikis2008}
{Deveikis}, V., {Narbutis}, D., {Stonkut{\.e}}, R., {Brid{\v z}ius}, A., \&
  {Vansevi{\v c}ius}, V. 2008, Baltic Astronomy, 17, 351

\bibitem[{{Dowell} {et~al.}(2008){Dowell}, {Buckalew}, \& {Tan}}]{Dowell2008}
{Dowell}, J.~D., {Buckalew}, B.~A., \& {Tan}, J.~C. 2008, \aj, 135, 823

\bibitem[{{Elmegreen} \& {Hunter}(2010)}]{Elmegreen2010}
{Elmegreen}, B.~G., \& {Hunter}, D.~A. 2010, \apj, 712, 604

\bibitem[{{Fall} {et~al.}(2005){Fall}, {Chandar}, \& {Whitmore}}]{Fall2005}
{Fall}, S.~M., {Chandar}, R., \& {Whitmore}, B.~C. 2005, \apjl, 631, L133

\bibitem[{{Fall} {et~al.}(2009){Fall}, {Chandar}, \& {Whitmore}}]{Fall2009}
---. 2009, \apj, 704, 453

\bibitem[{{Fioc} \& {Rocca-Volmerange}(1997)}]{Fioc1997}
{Fioc}, M., \& {Rocca-Volmerange}, B. 1997, \aap, 326, 950

\bibitem[{{Fouesneau} \& {Lan{\c c}on}(2010)}]{Fouesneau2010}
{Fouesneau}, M., \& {Lan{\c c}on}, A. 2010, \aap, 521, A22

\bibitem[{{Gil de Paz} {et~al.}(2007){Gil de Paz}, {Boissier}, {Madore},
  {Seibert}, {Joe}, {Boselli}, {Wyder}, {Thilker}, {Bianchi}, {Rey}, {Rich},
  {Barlow}, {Conrow}, {Forster}, {Friedman}, {Martin}, {Morrissey}, {Neff},
  {Schiminovich}, {Small}, {Donas}, {Heckman}, {Lee}, {Milliard}, {Szalay}, \&
  {Yi}}]{GildePaz2007}
{Gil de Paz}, A., {et~al.} 2007, \apjs, 173, 185

\bibitem[{{Girardi} \& {Bica}(1993)}]{Girardi1993}
{Girardi}, L., \& {Bica}, E. 1993, \aap, 274, 279

\bibitem[{{Groenewegen} \& {de Jong}(1993)}]{Groenewegen1993}
{Groenewegen}, M.~A.~T., \& {de Jong}, T. 1993, \aap, 267, 410

\bibitem[{{Hunter} {et~al.}(2003){Hunter}, {Elmegreen}, {Dupuy}, \&
  {Mortonson}}]{Hunter2003}
{Hunter}, D.~A., {Elmegreen}, B.~G., {Dupuy}, T.~J., \& {Mortonson}, M. 2003,
  \aj, 126, 1836

\bibitem[{{Johnson} {et~al.}(2011){Johnson}, {Seth}, {Dalcanton}, {Caldwell},
  {Gouliermis}, {Hodge}, {Larsen}, {Olsen}, {San Roman}, {Sarajedini}, {Weisz},
  \& {the PHAT Collaboration}}]{Johnson2011}
{Johnson}, L.~C., {et~al.} 2011, arXiv:1107.2668

\bibitem[{{Kroupa} \& {Boily}(2002)}]{Kroupa2002}
{Kroupa}, P., \& {Boily}, C.~M. 2002, \mnras, 336, 1188

\bibitem[{{Kroupa} {et~al.}(1993){Kroupa}, {Tout}, \& {Gilmore}}]{Kroupa1993}
{Kroupa}, P., {Tout}, C.~A., \& {Gilmore}, G. 1993, \mnras, 262, 545

\bibitem[{{Lamers} {et~al.}(2005){Lamers}, {Gieles}, \& {Portegies
  Zwart}}]{Lamers2005}
{Lamers}, H.~J.~G.~L.~M., {Gieles}, M., \& {Portegies Zwart}, S.~F. 2005, \aap,
  429, 173

\bibitem[{{Lan{\c c}on} \& {Mouhcine}(2000)}]{Lancon2000}
{Lan{\c c}on}, A., \& {Mouhcine}, M. 2000, in Astronomical Society of the
  Pacific Conference Series, Vol. 211, Massive Stellar Clusters, ed. {A.~Lan{\c
  c}on \& C.~M.~Boily}, 34

\bibitem[{{Larsen}(2009)}]{Larsen2009}
{Larsen}, S.~S. 2009, \aap, 494, 539

\bibitem[{{Larsen} \& {Richtler}(2000)}]{Larsen2000}
{Larsen}, S.~S., \& {Richtler}, T. 2000, \aap, 354, 836

\bibitem[{{Leitherer} \& {Chen}(2009)}]{Leitherer2009}
{Leitherer}, C., \& {Chen}, J. 2009, New Astron., 14, 356

\bibitem[{{Lejeune} {et~al.}(1997){Lejeune}, {Cuisinier}, \&
  {Buser}}]{Lejeune1997}
{Lejeune}, T., {Cuisinier}, F., \& {Buser}, R. 1997, \aaps, 125, 229

\bibitem[{{Ma{\'{\i}}z Apell{\'a}niz}(2009)}]{Maiz2009}
{Ma{\'{\i}}z Apell{\'a}niz}, J. 2009, \apss, 324, 95

\bibitem[{{Parmentier} \& {de Grijs}(2008)}]{Parmentier2008}
{Parmentier}, G., \& {de Grijs}, R. 2008, \mnras, 383, 1103

\bibitem[{{Piskunov} {et~al.}(2009){Piskunov}, {Kharchenko}, {Schilbach},
  {R{\"o}ser}, {Scholz}, \& {Zinnecker}}]{Piskunov2009}
{Piskunov}, A.~E., {Kharchenko}, N.~V., {Schilbach}, E., {R{\"o}ser}, S.,
  {Scholz}, R., \& {Zinnecker}, H. 2009, \aap, 507, L5

\bibitem[{{Popescu} \& {Hanson}(2010)}]{Popescu2010b}
{Popescu}, B., \& {Hanson}, M.~M. 2010, \apj, 724, 296

\bibitem[{{Schlegel} {et~al.}(1998){Schlegel}, {Finkbeiner}, \&
  {Davis}}]{Schlegel1998}
{Schlegel}, D.~J., {Finkbeiner}, D.~P., \& {Davis}, M. 1998, \apj, 500, 525

\bibitem[{{Searle} {et~al.}(1980){Searle}, {Wilkinson}, \&
  {Bagnuolo}}]{Searle1980}
{Searle}, L., {Wilkinson}, A., \& {Bagnuolo}, W.~G. 1980, \apj, 239, 803

\bibitem[{{Thim} {et~al.}(2003){Thim}, {Tammann}, {Saha}, {Dolphin}, {Sandage},
  {Tolstoy}, \& {Labhardt}}]{Thim2003}
{Thim}, F., {Tammann}, G.~A., {Saha}, A., {Dolphin}, A., {Sandage}, A.,
  {Tolstoy}, E., \& {Labhardt}, L. 2003, \apj, 590, 256

\bibitem[{{Whitmore} {et~al.}(2007){Whitmore}, {Chandar}, \&
  {Fall}}]{Whitmore2007}
{Whitmore}, B.~C., {Chandar}, R., \& {Fall}, S.~M. 2007, \aj, 133, 1067

\bibitem[{{Whitmore} \& {Zhang}(2002)}]{Whitmore2002}
{Whitmore}, B.~C., \& {Zhang}, Q. 2002, \aj, 124, 1418

\bibitem[{{Whitmore} {et~al.}(2011){Whitmore}, {Chandar}, {Kim}, {Kaleida},
  {Mutchler}, {Stankiewicz}, {Calzetti}, {Saha}, {O'Connell}, {Balick}, {Bond},
  {Carollo}, {Disney}, {Dopita}, {Frogel}, {Hall}, {Holtzman}, {Kimble},
  {McCarthy}, {Paresce}, {Silk}, {Trauger}, {Walker}, {Windhorst}, \&
  {Young}}]{Whitmore2011}
{Whitmore}, B.~C., {et~al.} 2011, \apj, 729, 78

\bibitem[{{Wofford} {et~al.}(2011){Wofford}, {Leitherer}, \&
  {Chandar}}]{Wofford2011}
{Wofford}, A., {Leitherer}, C., \& {Chandar}, R. 2011, \apj, 727, 100

\end{thebibliography}


\appendix
\section{Appendix A: More details on the comparison between the continuous and the
stochastic analysis}

This appendix provides some additional details about the comparison described in
Sect.\,\ref{sec:ContVsStoch}.

\begin{figure}
  \begin{center}
        \includegraphics[width=\columnwidth]{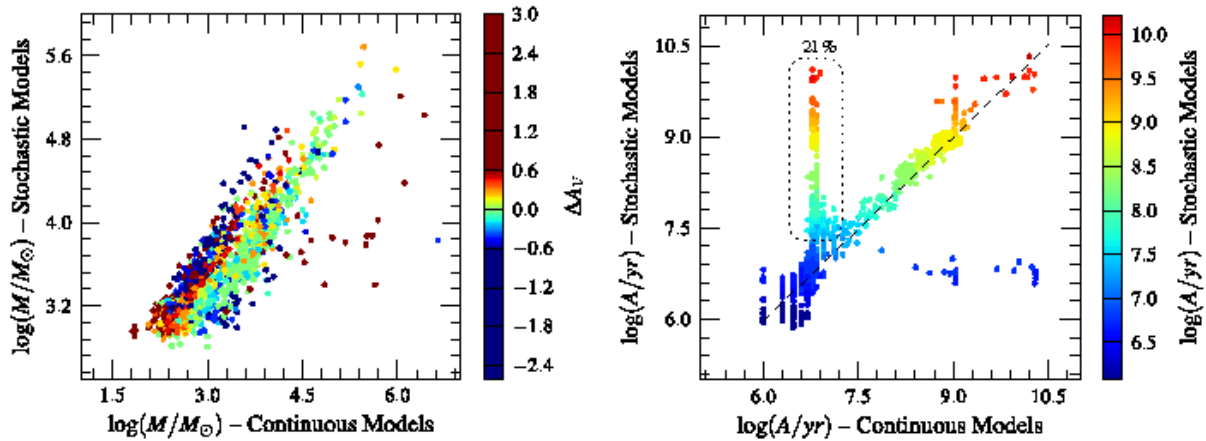}
  \end{center}
        \caption{Comparison of the ages and masses obtained when analysing the
        M\,83 cluster sample with ``continuous'' models on one hand and with
        ``stochastic'' models on the other. The dashed line 
         highlights the one-to-one relation,
	and the dotted box the vertical sequence mentioned in the text. 
\label{fig:CvsStoch}}
\end{figure}

In Fig.\,\ref{fig:CvsStoch}, we provide the direct cluster-by-cluster comparison
between the ages and masses derived { on one hand 
from an analysis with {\em stochastic} models, 
on the other hand from the best-$\chi^2$ fit to {\em continuous} models
(both based on the population synthesis code {\sc P\'egase}, and on 
UBVI and H$\alpha$ photometry). 
It is essential to note that in this section (as in most of the paper)
the parameters we assign in the stochastic context are {\em not} those 
of the single Monte-Carlo model that provides the best $\chi^2$ fit,
but those that maximize the posterior probability distribution of 
the cluster observations.} 
$72$\,\% of the clusters have ages within $0.3$\,dex
of each other, $28$\,\% have ages more than $0.3$\,dex apart. For the clusters
with similar ages in both approaches, 
the residuals have a bell-shaped distribution with a FWHM of
$0.3$\,dex. Most of the clusters with highly method-dependent ages lie on a
vertical sequence in the { second} panel of Fig.\,\ref{fig:CvsStoch} 
($21$\,\% of
all clusters, see the box on the figure).  They are assigned young ages with
continuous models, while the analysis based on stochastic models leads
to a range of older ages. The young ages of the continuous analysis come with
high values of the extinction parameter (because young cluster models are
intrinsically blue) and low masses (because young cluster models are
intrinsically bright), while the older ages of the stochastic analysis
are associated with low extinction parameters, and masses that are typically
$0.8$\,dex higher. This is seen in the { first}
panel of the figure as a one-sided
broadening of the one-to-one relation towards higher ``stochastic'' masses (dark
blue symbols).  Because this happens at all masses, the consequence for the mass
distribution is a slight offset in $\log(M/\msun)$ with hardly any change in the
slope of a fitted power law. The consequence for the age distribution is a
change in the strength of the peak of young clusters ($<10^7\yr$), and a
redistribution of these clusters over other ages in a way that again has little
effect on the slope of a fitted power law.  Note that for comparison with other
work in the literature one also has to keep in mind differences in the
population synthesis assumptions of various authors (see e.g.  Appendix B).

\begin{figure}
  \begin{center}
        \includegraphics[clip=,width=\columnwidth]{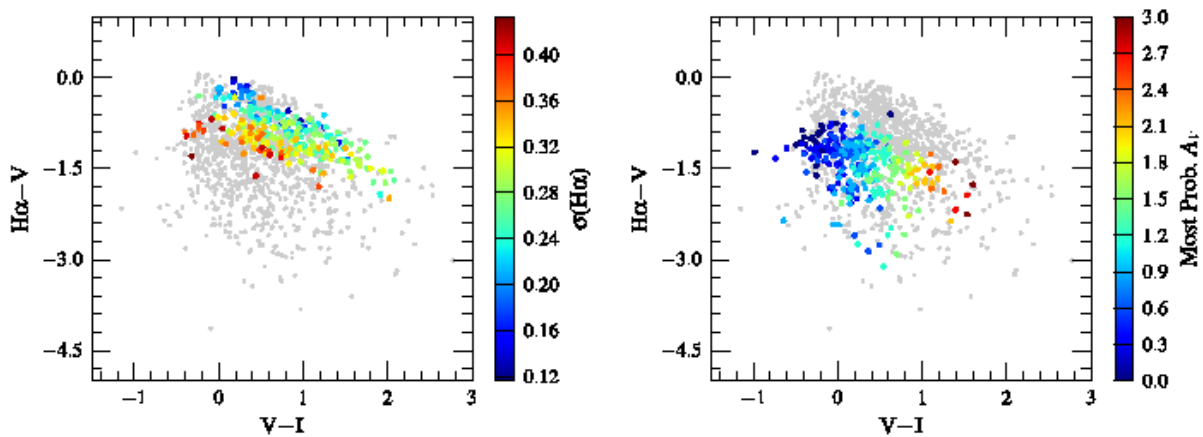}
  \end{center}
        \caption{Color-color plots similar to Fig.\,\ref{fig:halpha_colcoldiag},
        but with specific objects highlighted. In the first panel,
        the { colored} symbols identify the clusters 
        that populate the main vertical
        structure of the { second} panel of 
        Fig.\,\ref{fig:CvsStoch} (i.e.  the ``deviant clusters").
        The color code shows the observational uncertainty on the 
        clusters' H$\alpha$ measurement (in magnitudes).
        In the second panel, the { colored} symbols highlight low
        mass clusters whose ages are similar in the stochastic and in
        the continuous contexts, but whose masses differ. Here 
        the color code shows the associated A$_V$. Both subsets
        are discussed in the text. 
\label{fig:CvsStoch_interpret}}
\end{figure}

\medskip

The remainder of this appendix is provided for readers particularly interested
in the subtleties of the behaviour { of our analysis with
stochastic models. It gives further insight into the roles played by the prior
distributions of cluster ages and masses, and by the observational error bars.}

Color-color and color-magnitude plots can be used to understand the origin of
the vertical sequence in the age diagram just described. For brevity we will
call these particular clusters the ``deviant clusters'' in this paragraph. We
looked at many color combinations and confirmed that the deviant clusters are
located in a region of broad band color space where young reddened models and
old models overlap. The projection we found most useful for this discussion is
(\halpha$-$V) vs (V$-$I).  The deviant clusters are highlighted in this diagram
in the first panel of Fig.\,\ref{fig:CvsStoch_interpret}, which must be compared
to Fig.\,\ref{fig:halpha_colcoldiag}. About one half of the deviant clusters 
(shown in blue and cyan) lie along the line where most 
of the old and intermediate age synthetic clusters congregate,
near the upper envelope of the total cluster sample in this diagram.
The fact that they are assigned intermediate and old ages in the stochastic
analysis is an immediate consequence of the high density of intermediate
and old model clusters in that area.  The fact that some are assigned 
very young ages in the continuous analysis appears as a consequence 
of the ``single best fit'' approach of that continuous analysis:
by chance the model closest to them happens to be a young one, but a 
small shift in the observed colors (well within the photometric error bars)
could change the derived age significantly since young reddened models 
and older dust-free ones are found side by side.
The second half of the deviant
clusters (shown in yellow and red) are located in a region where only young 
reddened model clusters exist. These objects however {\em all} have errors 
larger than $0.25$ magnitudes on the F657N
measurement (and the errors increase for deviant clusters located 
further away from the upper envelope of the data points). 
{ For these clusters we see the effects of the observational errors clearly.
The multi-dimensional $2\,\sigma$ error boxes around each of them 
reach well into the above-mentioned region of color-color space 
where the density of intermediate age and old clusters is high.  
Models outside such an error box contribute negligibly 
to the posterior probability distribution, but all models inside 
may contribute significantly (Bayes' theorem).  
Because the error boxes contain overwhelmingly more intermediate 
and old models than young reddened models, our analysis favours 
the older solution.} For clusters at the same location
in color-color space but with much smaller error bars on \halpha, 
young ages are returned. Note that the number of clusters 
concerned by this discussion is too small to change 
any of the main results discussed in this paper.

A second more subtle effect contributes to the one-sided broadening of the mass
vs. mass plot, this time only in the regime of small masses ($\log(M/\msun) <
3.5$ in the continuous analysis -- red  symbols in the { first} panel of
Fig.\,\ref{fig:CvsStoch}).  These objects have identical ages with the
continuous and with the stochastic analysis (frequently within $0.1$\,dex), but
the ``stochastic'' masses are larger than the ``continuous'' masses by
$0.3-0.6$\,dex. Their location in the most useful color-color diagram is
highlighted in the second panel of Fig.\,\ref{fig:CvsStoch_interpret}.  We see
in Fig.\,\ref{fig:halpha_colcoldiag} that the vast majority of the
young stochastic cluster models lie {\em not} along the line of models with a
continuously sampled IMF, but rather on a curved ``plume'' at bluer colors. This
would be the case for any prior mass distributions that has more low mass
clusters than high mass clusters.  As a consequence, { the analysis
of the posterior probability distribution}
favors the models in this bluer ``plume'' and assigns a correspondingly larger
extinction. The age changes little when compared to the analysis with a
continuously sampled IMF, but the larger extinction must be compensated for with
a somewhat larger mass. This effect concerns about $20$\,\% of the clusters. It
has a noticeable effect on the mass distribution derived from stochastic models
only below $\log(M/\msun)=3.6$.

\section{Appendix B : comparison with the ages and masses of Chandar et al. 2010}

\citet{Chandar2010} use continuous population synthesis models. Figure 10 of
their article shows the age-mass distributions they obtain when assuming solar
and twice solar metallicity. They are qualitatively very similar to the
distribution shown here in Fig.\,\ref{fig:StdAgeMasDistrib}.

The differences are due in part to the underlying population synthesis
assumptions, in part to the analysis method. \citet{Chandar2010} use models from
Bruzual \& Charlot (2009) with a Chabrier IMF, and they account for hydrogen
line emission with an escape fraction of $0.4$ (the value that, in their
analysis, provides best fit qualities on average). The fit for age and
extinction is done using magnitudes, and the mass is then determined from the V
band absolute magnitude.  As a consequence of using magnitudes in their case,
fluxes in our case, the different photometric bands are given different weights
in both studies, and the reduced $\chi^2$ values cannot be compared directly
(except very grossly: in both cases a $\chi^2$ value above a few indicates a
poor fit).

\begin{figure}
        \begin{center}
                \includegraphics[]{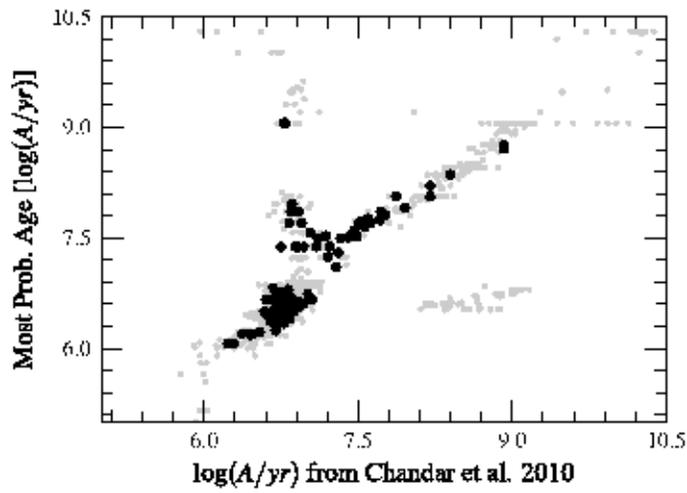}
        \end{center}
	\caption{Ages derived from with {\it continuous} models by
	\citet{Chandar2010} [$x$-axis] and in this paper [$y$-axis]. Large black
	dots highlight the $114$ clusters for which the error in all filters are
	smaller than $0.13$ mag. 
        \label{fig:bc09pegase}
        }
\end{figure}

Figure\,\ref{fig:bc09pegase} directly compares ages retrieved by
\citet{Chandar2010} and ages we obtain in this present paper.  Ages are similar
for $85$\,\% of the clusters (and for $95$\,\% of the clusters with small
observational errors). These objects are on a curved line that deviates slightly
from the one-to-one diagonal, so that strictly speaking only 75\,\% of all the
clusters have ages that agree to within 0.3\,dex with both methods.  Differences
one could call ``catastrophic'' occur for $15$\,\% of the clusters, and are not
restricted to observations with large error bars. These differences depend
rather sensitively on the way nebular emission is accounted for in the models
(the vertical branch of ``catastrophic errors'' disappears and the horizontal
one gains clusters when we remove the nebular emission from the continuous
models used here).  
They also depend on the range over which the extinction parameter is
allowed to vary.

\end{document}